\documentclass[format=acmsmall, review=false, screen=true]{acmart}

\usepackage{booktabs} 

\usepackage{graphicx}
\usepackage{amsmath}
\usepackage{amsthm}
\usepackage[noend]{algpseudocode}
\usepackage{epstopdf}
\usepackage{algorithm}
\usepackage{color}
\usepackage{subcaption}
\usepackage{multirow}
\usepackage{framed}
\usepackage[T1]{fontenc}
\usepackage[latin1]{inputenc}
\usepackage{tabularx}
\usepackage{tikz}
\usetikzlibrary{matrix}
\usepackage{bm}
\newcommand{\hide}[1]{\ifthenelse{\boolean{false}}{#1}{}}

\usepackage{gensymb}
\usepackage{latexsym,epsfig,color,verbatim}
\usepackage{pifont}
\usepackage{slashbox}
\usepackage{bbm}
\newtheorem{remark}{\bf Remark}

\newcommand{\black}[1]{\textcolor{black}{#1}}

\newtheorem{theorem}{{\bf Theorem}}
\newtheorem{lemma}{{\bf Lemma}}
\newtheorem{claim}{{\bf Claim}}

\newtheorem{corollary}{{\bf Corollary}}

\newtheorem{defn}{Definition}

\newcommand{\barr}{\begin{array}}
\newcommand{\earr}{\end{array}}

\newcommand{\benum}{\begin{enumerate}}
\newcommand{\eenum}{\end{enumerate}}

\newcommand{\bit}{\begin{itemize}}
\newcommand{\eit}{\end{itemize}}

\newcommand{\bdes}{\begin{description}}
\newcommand{\edes}{\end{description}}

\newcommand{\bfig}{\begin{figure}}
\newcommand{\efig}{\end{figure}}

\newcommand{\bemq}{\begin{quote} \begin{em}}
\newcommand{\eemq}{\end{em} \end{quote}}

\newcommand{\ul}{\underline}

\newcommand{\bt}{\begin{theorem}}
\newcommand{\bl}{\begin{lemma}}
\newcommand{\bc}{\begin{claim}}
\newcommand{\bp}{\begin{Proposition}}
\newcommand{\bcoro}{\begin{corollary}}
\newcommand{\bres}{\begin{Result}}
\newcommand{\brem}{\begin{Remark}}
\newcommand{\et}{\end{theorem}}
\newcommand{\el}{\end{lemma}}
\newcommand{\ec}{\end{claim}}
\newcommand{\ep}{\end{Proposition}}
\newcommand{\ecoro}{\end{corollary}}
\newcommand{\eres}{\end{Result}}
\newcommand{\erem}{\end{Remark}}

\newcommand{\beq}{\begin{equation}}
\newcommand{\eeq}{\end{equation}}

\newcommand{\UN}[1]{{\mathcal{V}}^{(N)}_{k}}
\newcommand{\norm}[1]{\|{#1}\|}
\newcommand{\abs}[1]{\left \vert {#1} \right \vert}

\newcommand{\mb}[1]{\mathbb{#1}}

\begin{document}
\title{Proximity Based Load Balancing Policies on Graphs: A Simulation Study}

\author{Nitish K. Panigrahy}
\affiliation{
  \institution{College of Information and Computer Sciences, UMass Amherst}
  \state{Amherst, MA 01003, USA} 
}
\email{nitish@cs.umass.edu}

\author{Thirupathaiah Vasantam}
\affiliation{
  \institution{College of Information and Computer Sciences, UMass Amherst }
  \state{Amherst, MA 01003, USA} 
}
\email{tvasantam@cs.umass.edu}

\author{Prithwish Basu}
\affiliation{
  \institution{Raytheon BBN Technologies}
  \state{Cambridge, MA 02138} 
}
\email{prithwish.basu@raytheon.com}

\author{Don Towsley}
\affiliation{
  \institution{College of Information and Computer Sciences, UMass Amherst}
  \state{Amherst, MA 01003, USA} 
}
\email{towsley@cs.umass.edu}


\begin{abstract}

Distributed load balancing is the act of allocating jobs among a set of servers as evenly as possible. There are mainly two versions of the load balancing problem that have been studied in the literature: static and dynamic. The static interpretation leads to formulating the load balancing problem as a classical balls and bins problem with jobs (balls) never leaving the system and accumulating at the servers (bins) whereas the dynamic setting deals with the case when jobs arrive and leave the system after service completion. While most of the previous work in the static setting focus on studying the maximum number of jobs allocated to a server or \emph{maximum load}, little importance has been given to the distributional aspect of load associated with the allocation policies. Also it is desirable to study load balancing models that incorporates the \emph{access latencies} (time that elapses from when the job is created until it is completed) or \emph{communication cost} associated with of jobs since such scenarios occur widely in practice. 

This paper designs and evaluates server proximity aware job allocation policies for treating load balancing problems with a goal to reduce the communication cost associated with the jobs. We consider a class of proximity aware Power of Two (POT) choice based assignment policies for allocating jobs to servers, where servers are interconnected as an $n$-vertex graph $G(V, E)$. For the static version, we assume each job arrives at one of the servers, $u,$ chosen uniformly at random from the vertex set $V.$ For the dynamic setting, we assume $G$ to be a circular graph and job arrival process at each server is described by a Poisson point process with the job service time exponentially distributed. For both settings, we then assign each job to the server with minimum load among servers $u$ and $v$ where $v$ is chosen according to one of the following two policies: (i) Unif-POT($k$): Sample a server $v$  uniformly at random from $k$-hop neighborhood of $u$ (ii) InvSq-POT($k$): Sample a server $v$  from $k$-hop neighborhood of $u$ with probability proportional to the inverse square of the distance between $u$ and $v$. 

An extensive simulation over a wide range of topologies validate the efficacy of both the proximity aware load balancing policies. Our simulation results for static systems show that both the policies consistently produce a load distribution which is much similar to that of a classical proximity oblivious POT policy. Depending on topology, we observe the total variation distance to be of the order of $0.2\% - 8\%$ for both the policies while achieving a $8\%-99\%$ decrease in communication cost as compared to classical POT policy. More precisely,  we observe that InvSq-POT($k$) achieves the best of both the worlds, i.e. it is better than UnifPOT($k$) in terms of communication cost but only slightly worse than classical POT in terms of load distributions. For dynamic load balancing system, although we observe a drop in performance,  we also get a significant drop in the communication cost associated with the two policies compared to classical POT over a ring topology. 
\end{abstract}

\maketitle

\section{Introduction}\label{sec:intro}

The past few years have witnessed an increased interest in the use of large-scale parallel and distributed systems for database and commercial applications. A massively parallel processing system consists of a large number of small server nodes joined together via a network. An important design goal in such a system is to distribute service requests or jobs among servers or distributed resources as evenly as possible. While the optimal server selection problem can be solved centrally, due to scalability concerns, it is often preferred to adopt distributed randomized load balancing strategies to distribute these jobs among servers. These strategies have proven to provide good performance in terms of reducing collisions, waiting times and often achieving high resource utilization levels. 

There are mainly two versions of a load balancing problem: static and dynamic. The case when jobs never depart and just accumulate is known as the static load balancing system \cite{Azar99}. The static interpretation leads to formulating the load balancing as a  classical balls and bins problem. Here, jobs are mapped to balls and servers are mapped to bins. In the classical balls-and-bins model, $m$ balls are placed sequentially into $n$ bins. Each ball uniformly at random, from the entire set of bins,  selects a bin also known as the origin bin (server) for the corresponding ball. The ball then samples $d-1$ bins uniformly at random from the $n-1$ bins and is allocated to the bin with the least load among the sampled and the origin bin with ties broken arbitrarily. For the case  when $d = 1$ and $m = n,$ this assignment policy results in a maximum load of $O(\log n/\log \log n)$ with high probability (w.h.p). When $d = 2$, the maximum load is $O(\log \log n)$ w.h.p. \cite{Azar99} and the policy is widely known as \emph{Power of Two} (POT) choices. Note that there is an exponential improvement in the performance from $d = 1$ to $d = 2$. This improvement in maximum load behavior is often called \emph{POT benefits} \cite{Panigrahy2020}. 

One of the fundamental issues with the POT policy lies in the fact that it involves fetching load information from any $d-1$  servers. This may induce high communication cost, particularly for massively parallel and distributed systems spanning over a wide geographic area. For example, the cost of accessing two servers that are far away may be larger than that of accessing two nearby servers. Since a POT policy is oblivious to these server proximities associated with the sampling process, the following natural question arises.\\
\begin{quotation}
{\it How should we design proximity aware load balancing policies that also reduce overall job-to-server communication cost?}
\end{quotation}
One simple solution to the proximity problem is to sample $d-1$ servers in a correlated way. Such correlations can be captured by representing the set of servers as vertices of an arbitrary graph $G(V,E).$ In the arbitrary graph model, when a job arrives at a server $u$, it is assigned to the server with the least load among server $u$ and $d-1$ servers sampled uniformly at random from its one hop neighborhood in $G.$ Jobs arriving to a server with less than $d-1$ neighbors are assigned in an arbitrary fashion among $u$ and its neighbors without probing the load values. 

In order to make the arbitrary graph model more tractable for theoretical analysis, many simplified assumptions have been made over the graph structure. For example,  Kenthapadi et al. \cite{Kenthapadi2005} studied the scaling  of maximum load for the case $d = 2$ by allowing $G$ to be regular or almost regular with degree $n^\epsilon$. However, in practice, real world networks are highly irregular and some are deterministic with fixed degrees. Moreover, state-of-the-art lacks a comprehensive study on characterizing the communication cost associated with the arbitrary graph model. More often previous work only considers developing theoretical framework to characterize the scaling of the maximum load behavior and ignores distributional aspect of load. These bring us down to the following research questions.

\begin{enumerate}
\item How should one evaluate the performance of proximity aware load balancing policies for more general non-regular graph models,  such as random, scale free or spatial graph structures? \\
\item What is a good performance metric to characterize the communication cost associated with the arbitrary graph model?\\
\item What is the effect on load distribution and communication cost if $d-1$ servers are sampled from $k$-hop instead of one hop neighborhood of origin server with $k \ge 2$?\\
\item How close is the performance of a proximity based POT policy to POT policy with respect to load distribution instead of maximum load metric?
\end{enumerate}

The primary motivation behind this work is to address these research questions. The key challenge in developing theoretical frameworks to answer these questions is that the notion of neighborhood for each job heavily depends on a particular choice of graph topology. Unlike POT, server states are not exchangeable  and techniques like witness tree methods \cite{Kenthapadi2005} are not applicable. Thus even asymptotic results for proximity aware load balancing policies on an arbitrary graph are scarce. There is little hope of analyzing the arbitrary graph model in this generality and generating analytical insights seems difficult to achieve. For that reason we investigate this model through detailed and extensive computer simulations across a variety of graph topologies in Section \ref{sec:bandb}. 

We first define a proximity based policy: \emph{Unif-POT($k$)} considers the $k$-hop neighborhood around the origin server as follows. For each job, we sample a server $v$  uniformly at random from the set of $k$-hop neighborhood of the origin server $u.$ Since a POT policy optimally balances load (by stochastic majorization argument), we compare the load distribution of Unif-POT($k$) policy to that of POT policy. Surprisingly, the \emph{total variation distance}, a metric to determine closeness of two probability distributions, between Unif-POT($k$) and POT is very close to zero across a wide range of both finite degree deterministic and random network topologies. 
We also observe a drastic reduction in the average number of hops between the origin server and the allocated server for a job, also known as \emph{average request distance}, for Unif-POT($k$)  policy as compared to POT. 

Coming back to designing proximity aware load balancing policies, another strategy is to sample a candidate set of $d-1$ servers in a non-uniform communication cost  dependent manner. Recently, Panigrahy et al. \cite{Panigrahy2020} proposed a new class of POT based geometric load balancing policies for allocating jobs to servers, where servers are located on a two-dimensional Euclidean plane. Authors in \cite{Panigrahy2020} sample two servers each with probability proportional to $1/x^2$ where $x$ is the Euclidean distance between the location of the job and the concerned server. The job is then allocated to the server with least load. However, the simulation setup in \cite{Panigrahy2020} is cursory lacking a comprehensive study of the proposed policy across different system parameters, even in the geometric setting.

\black{One of our contributions in this paper is to rigorously study the non-uniform sampling based proximity  aware load balancing policies in a general graph setting.} We define such a load balancing policy: \emph{InvSq-POT($k$)} on arbitrary graphs as follows. For each job, we sample a server $v$  from the $k$-hop neighborhood of origin server $u$ with probability proportional to the inverse square of the shortest path distance between $u$ and $v$. Again through extensive simulations we verify that such a simple modification in the sampling technique, produces load distribution behavior very similar to that of POT policy while drastically reducing the average request distance across a variety of network topologies. For certain random network topologies, surprisingly enough, even a very local sampling, i.e., InvSq-POT($k$) with $k = O(1)$ achieves a load balancing performance almost similar to that of classical POT policy.

POT load  benefits are also observed in dynamic load balancing systems, i.e. in systems where incoming jobs are assigned to one of the servers according to POT policy and then leave the system after getting served according to a service discipline. In particular, the tail of the queue length distribution decreases doubly exponentially when $d=2$ as compared to single exponentially when $d=1$  \cite{mitzenmacher96}. 

Recently, the arbitrary graph model with one hop neighborhood was studied by Budhiraja et al. \cite{Budhiraja19} for dynamic systems . However, some of the previous research questions, particularly questions $(2)-(4)$, are still open for the dynamic systems as well. Again, for proximity based policies on arbitrary graphs, often  servers are not exchangeable and the standard  mean-field techniques do not apply as locations of the servers also play a role in job assignment. Since the analysis is challenging, we draw insights into cost vs performance tradeoff through extensive simulations for ring topology.

Our contributions are summarized below.
\begin{itemize}
\item {\bf Results for static load balancing systems} 
\begin{enumerate}
\item We perform extensive simulations to evaluate the performance of proximity based load balancing policies: Unif-POT($k$) and InvSq-POT($k$) for static load balancing systems across a wide range of network topologies such as: deterministic, random, scale-free and spatial networks.
\item We achieve a total variation distance as low as $0.2\%-0.5\%$ between load distributions of classical POT and proximity based policies.
\item  We achieve a significant reduction in communication cost on the order of $20\%-99\%$ for proximity based policies as compared to classical POT policy.
\item We observe that InvSq-POT($k$) with $k = O(\log n)$ achieves the best of both the worlds, i.e. it is better than UnifPOT($k$) in terms of communication cost but only slightly worse than classical POT in terms of load distributions.  To our surprise, even a very local sampling, i.e., InvSq-POT($k$) with $k = O(1)$ achieves a load balancing performance almost similar to that of classical POT policy for certain random network topologies.
\end{enumerate}
\item {\bf Results for dynamic load balancing systems}
\begin{enumerate}
\item We obtained simulation results on proximity based load balancing policies for dynamic load balancing systems over a ring topology. Although we observe a drop in performance,  we also get a significant drop in communication cost associated with both proximity based policies.
\item We achieved a reduction in communication cost of the order of at least $25\%$ for InvSq-POT($k$) policy as compared to Unif-POT($k$) and classical POT policy.
\item For fixed $k,$ we observe that the load distributions for both proximity based policies converge to a limit as the number of servers $n$ becomes large.
\end{enumerate}
\end{itemize}

The rest of this paper is organized as follows. The next section contains some related literature. In Section \ref{sec:model} we discuss some technical preliminaries. Section \ref{sec:prox} introduces two proximity based load balancing policies for graphs. Through extensive simulations, we evaluate both the proximity based policies for static system in Section \ref{sec:bandb}. We study the dynamic load balancing system in Section \ref{sec:dyn}. Finally, the conclusion of this work and potential future work are given in Section \ref{sec:con}.

\section{Related Work}\label{sec:reltwrk}
The load balancing problem can be categorized into two versions: static and dynamic. We first discuss the state-of-the-art related to the static version as below.
\subsection{Static Load Balancing}

Many previous works \cite{Adler1998, mitzenmacher96} have developed simple and efficient load balancing algorithms in the static setting. The widely acclaimed \emph{Power of Two} (POT) choices policy was first proposed by Azar et al. \cite{Azar99}. Further generalizations of POT policy to account for correlated and non-uniform sampling  strategies have been discussed in subsequent works \cite{Vocking2003, Berenbrink2006, byers04}. Load balancing on graphs was first proposed by Kenthapadi et al. \cite{Kenthapadi2005}, where the authors considered a model with bins interconnected as a $\Delta$-regular graph. Each ball then samples a random edge of the graph and gets allocated at one of its endpoints with smaller load. Godfrey et al. \cite{Godfrey2008} generalized the graph based model to balanced allocations on a hypergraph. Bringmann et al. \cite{Bringmann2016} studied a model where each ball picks a random bin and performs a local search from the bin to a bin with local minimum load and gets allocated to it. Pourmiri et al. \cite{Pourmiri2019} proposed algorithms for allocating balls to bins that are interconnected as a regular graph by performing a non-backtracking random walk from a chosen node. Recently, Panigrahy et al. \cite{Panigrahy2020} proposed a new class of POT based geometric load balancing policies for allocating balls to bins, where both balls and bins are located on a two-dimensional Euclidean plane. Authors in \cite{Panigrahy2020} sample two bins each with probability proportional to $1/x^2$ where $x$ is the Euclidean distance between the ball and the concerned bin. The ball is then allocated to the bin with least load.


\subsection{Dynamic Load Balancing}

In the dynamic setting, servers process assigned jobs according to the First-Come-First-Serve (FCFS) service discipline and a processed job departs the system up on completion of its service. It was shown in \cite{Vvedenskaya96,mitzenmacher96} that under the Power of $d$ ($d\geq 2$) policy, the stationary probability that a server has at least $i$ progressing jobs in the asymptotic regime when $n\to\infty$ is equal to $(\lambda/\mu)^{(d^{i}-1)/(d-1)}$, whereas it equals to $(\lambda/\mu)^i$ when $d=1$. This shows that the POT policy reduces the average delay significantly. However, if we consider the distance between the server where a job has emerged and the server where it is processed into implementation cost of the policy, which is true for many real world applications, then the POT policy has a drawback of high implementation cost. Therefore, it is of interest to study policies that consider both occupancy and location of a server in deciding the destination server for an arrival.

 Next, we discuss some existing relevant works that have studied load balancing on graphs. In \cite{Gast15}, a load balancing policy was investigated for symmetric graphs with degree $K$ in which jobs arrive at each server according to a Poisson process of rate $\lambda$ and a job that arrived at a server say $s$ is served at the shortest of servers $s$ and $m$, where $m$ is another server picked uniformly at random from the set of $K$ servers that are neighbors of server $s$. They were able to derive a set of evolution equations based on pair-wise approximations and the fixed-point of these equations was shown to approximate the stationary distribution of a server's state. They also showed that the mean-field approximations are not useful in this setting. 
 
 In \cite{Budhiraja19}, a similar model in which servers are located at the nodes of a deterministic graph $G_n$ was studied. They showed that if $d_{min}(G_n)\to\infty$ and 
 $\sup_{i\geq 1}\abs{[d_{min}(C_{i,n})/d_{max}(C_{i,n})]-1}\to 0$, where $d_{min}(G_n)$ indicates the minimum degree of $G_n$, $C_{i,n}$ is a connected component of the graph and $d_{max}(C_{i,n})$ denotes the maximum degree of a node in $C_{i,n}$, then the empirical process of occupancy converges to the same mean-field as in the case of POT policy. They also showed that for Erd\H{o}s-R\'{e}nyi graphs with average degree $D_N$, the empirical process of occupancy was shown to converge to the same mean-field limit as in the POT policy if $D_n/\ln(n)\to\infty$ as $n\to\infty$. Recently, in \cite{Debankur20}, a load balancing policy was studied for a bipartite graph in which task types are matched to servers which can serve them and each task type can be processed only at a small subset of servers. An incoming task is assigned to a server that has the shortest queue size among $d$ randomly chosen servers from the set of servers which can process it. Under the assumption that if a graph satisfies certain connectivity properties referred to as proportional sparsity, they showed that empirical occupancy process converges to the same mean-field limit as in the case of a complete bipartite graph. For random graphs that satisfy proportional sparsity, if the degree of a server is at most  $D_n$ satisfying $D_n\to\infty$ and $nD_n/M(n)\ln(n)\to\infty$ as $n\to\infty$, then the empirical occupancy process was shown to converge to the same mean-field limit as in the complete bipartite graph case. In this paper, we study ring topology which does not satisfy the assumptions considered in \cite{Debankur20, Budhiraja19}.

\section{Preliminaries}\label{sec:model}
In this section, we introduce our system model used in the rest of the paper. 
\subsection{Servers and Jobs}
\noindent {\bf Servers:} We assume servers in the network are nodes of a connected unweighted graph $G (V, E)$ with $|V| = n$ and  $E$ a  set of edges connecting the servers. We assume $G$ does not contain multiple edges. While we explore various random, deterministic and spatial graph structures for static load balancing systems, we perform simulations on a ring topology for dynamic systems.\\
\noindent {\bf Jobs:} For static load balancing systems, we assume that jobs arrive at one of the servers uniformly at random. Denote $u$ as the arrival (origin) server for a random job. We denote $J$ as the set of jobs and $m$ as the total number of jobs in the system. In our simulations, we mostly consider the case $m=n$. For dynamic systems, jobs arrive at each server according to a Poisson process of rate $\lambda$.  Then jobs are served according to FIFO service discipline. Service times are independent and identically distributed (IID)  exponential random variables with mean $\mu$. Upon an arrival at a server, the server samples another server from its $k$-hop neighborhood according to a predefined probability distribution. Once the load (or queue size) information of the sampled server is collected, the job is routed to the server with the least load (or queue size). Ties are broken randomly.

\subsection{Network Topologies}
We primarily consider the following graph topologies: deterministic, random and spatial graphs .
\subsubsection{Deterministic Graphs with fixed degrees}
Load balancing algorithms for certain fixed-degree deterministic graphs, in particular for ring topologies, has been studied in the past \cite{Gast15}, \cite{Turner1998} and have applications in many fields such as bike-sharing systems. We consider the following deterministic graphs with fixed degrees for our simulations.

\subsection*{Line Graph-{\it L($n$)}}
A Line graph L($n$) is a graph whose vertices $v_1, v_2, \cdots, v_n$ are connected with edges $(v_i, v_{i+1}),.i = 1,2,\cdots, n-1.$
\subsection*{Ring Graph-{\it R($n$)}}
The ring graph R($n$) on $n$ vertices can be viewed as having a vertex set $\{0, 1, \cdots, n-1\}$ corresponding to the integers modulo $n$ with edges $(i,i+1)$, modulo $n$.

\subsubsection{Random Graphs} Random networks with power law degree distributions are scale-free networks. Many complex networks, like the World Wide Web, can be modeled as scale free graphs. Similarly, the underlying topology of many peer-to-peer networks \cite{Cooper2007}  can be modeled as random regular graphs. Surprisingly, very few rigorous insights are known about load balancing policies on such random networks. In our simulations, we consider the following random networks.

\subsection*{Barabasi Albert Graph- {\it BA ($n,\alpha$)}:} A graph BA ($n,\alpha$) of $n$ nodes is grown by adding new nodes each with $\alpha$ edges attached to existing nodes with probability proportional to the node degree. This has been shown to yield power-law degree distribution.

\subsection*{Random Regular Graph- {\it RR ($n,\beta$)}:}  A $\beta$-regular graph RR ($n,\beta$) sampled from  the probability space of all $\beta$-regular graphs on $n$ vertices uniformly at random with $n\beta$ being even. For $\beta \ge 3,$, a random $\beta$-regular graph of large size is asymptotically almost surely $\beta$-connected. In all our simulations, we use $\beta \ge 3.$

\subsection*{Erdos-Renyi Graph- {\it ER ($n,\gamma$)}:} The ER ($n,\gamma$) graph is generated by choosing each of the $[n(n-1)]/2$ possible edges with probability $\gamma$. $\gamma = \log n /n$ is a sharp threshold for the connectedness of $G(n, \gamma).$ Also as $n\to\infty$, the probability that $G(n, \gamma)$ with $\gamma = 2\log n /n$ is connected, tends to $1$. In all of our simulations, we assume $\gamma \ge \log n /n.$

\subsubsection{Spatial  Graphs}
Many real-world networks exist in an euclidean space, thus come with a spatial embedding. For example, the communication network resulting from radio transmitters and wireless devices can be described by a random geometric graph \cite{Penrose2007}. Similarly, a line topology applies to vehicular wireless ad-hoc networks on a one-lane roadway~\cite{Ho11,Leung94}, where users are in vehicles submitting jobs and servers are attached to fixed infrastructure such as lamp posts. Such networks have a natural notion of distance, so it is important to take into account this geographical aspect when designing load balancing policies. Below we describe three different spatial graphs that we explored in our simulations.

\subsection*{2-D Random Geometric Graph-{\it RG ($n,r$)}}
A 2-D random geometric graph RG ($n,r$) is an undirected graph with $n$ nodes uniformly sampled from a $2$-dimensional euclidean space $[0,1)^2.$ Two vertices: $a,b \in V$ share an edge iff the euclidean distance between these two servers is less than $r$, excluding any loops. RG ($n,r$) possesses a sharp threshold for connectivity at $r \sim \sqrt{\log n/\pi n}.$ In all our simulations we consider $r \ge \sqrt{\log n/\pi n}.$
\subsection*{Spatial Line Graph-{\it SL($n, L_{max}$)}}
Locations of servers are uniformly sampled from a one-dimensional euclidean space $[0,L_{max}).$ We assume users that submit jobs are also placed on the same line uniformly at random.

\subsection*{Spatial Ring Graph-{\it SR($n, R$)}}
We assume servers and users that submit jobs are placed uniformly at random on a circle of radius $R$. 

\subsection{Network Attributes}
Next we define several network attributes that would be useful in analyzing the simulation results obtained for different load balancing policies later. We denote $\phi(u,v), u, v \in V$ as the shortest path distance between nodes $u$ and $v$ in the network.

\begin{defn}{$k$-hop Neighborhood:} The $k$-hop neighborhood of a node $u\in V$ is defined as
\begin{align}
\mathcal{N}_k(u) = \{w|1\le\phi(u,w)\le k\}.\nonumber
\end{align}
\end{defn}

\begin{defn}{Graph Density:} The graph density of an undirected graph $G(V, E)$ is
\begin{align}
\rho_G = \frac{|E|}{{n \choose 2}} = \frac{2|E|}{n(n-1)}.\nonumber
\end{align}
\end{defn}

\begin{defn}{Average Path Length:} The average path length of an undirected graph $G(V, E)$ is
\begin{align}
l_G = \frac{1}{n(n-1)}\sum\limits_{u \ne v}\phi(u,v).\nonumber
\end{align}
\end{defn}

\subsection{Performance Metrics}
To evaluate and characterize the performance of various load balancing policies, we define the performance metrics for static systems as follows.

Let $\pi: J \rightarrow V,$ denote a load balancing policy for assigning jobs to servers. Denote $x^\pi(t) = [x_i^\pi(t),i\in \{1,\cdots,m\}]$ as the state of the static system immediately after the $t^{th}$ job is assigned under policy $\pi$. Here $x_i^\pi(t)$ denotes the fraction of servers with exactly $i$ jobs immediately after $t^{th}$ job is assigned. Denote $x^\pi = x^\pi(m)$ as the load distribution under policy $\pi$ after all of  $m$ jobs are assigned.

\begin{defn}{Total Variation Distance:} The total variation distance between two  load distributions $x^{\pi_1}$ and $x^{\pi_2}$ is
\begin{align}
TV^{\pi_1\pi_2}= \frac{1}{2}\sum\limits_{i=1}^m | x^{\pi_1}_i - x^{\pi_2}_i|.\nonumber
\end{align}
\end{defn}
The closeness of two load distributions under two different policies can be measured by the total variation distance, i.e. smaller the total variation distance the closer the two distributions are to each other. In Section \ref{sec:bandb}, through extensive simulations, we show that various proximity based POT policies produce near zero total variation distance with respect to load distribution of the classical POT policy.

\begin{defn}{Average Request Distance:} The average request distance for policy $\pi$ is the average number of hops between the origin server and the allocated server for a random job under $\pi$, i.e.
\begin{align}
RD^\pi = \frac{1}{m}\sum\limits_{j\in J}\phi(u_j,\pi(j)).\nonumber
\end{align}
\end{defn}
Since POT is oblivious to inter server distances, $RD^{POT}$ is generally high as compared to other proximity based load balancing policies.

\section{Proximity Based POT Policies for General Graphs}\label{sec:prox}

\begin{figure}[!htbp]
\centering
\includegraphics[width=0.5\linewidth]{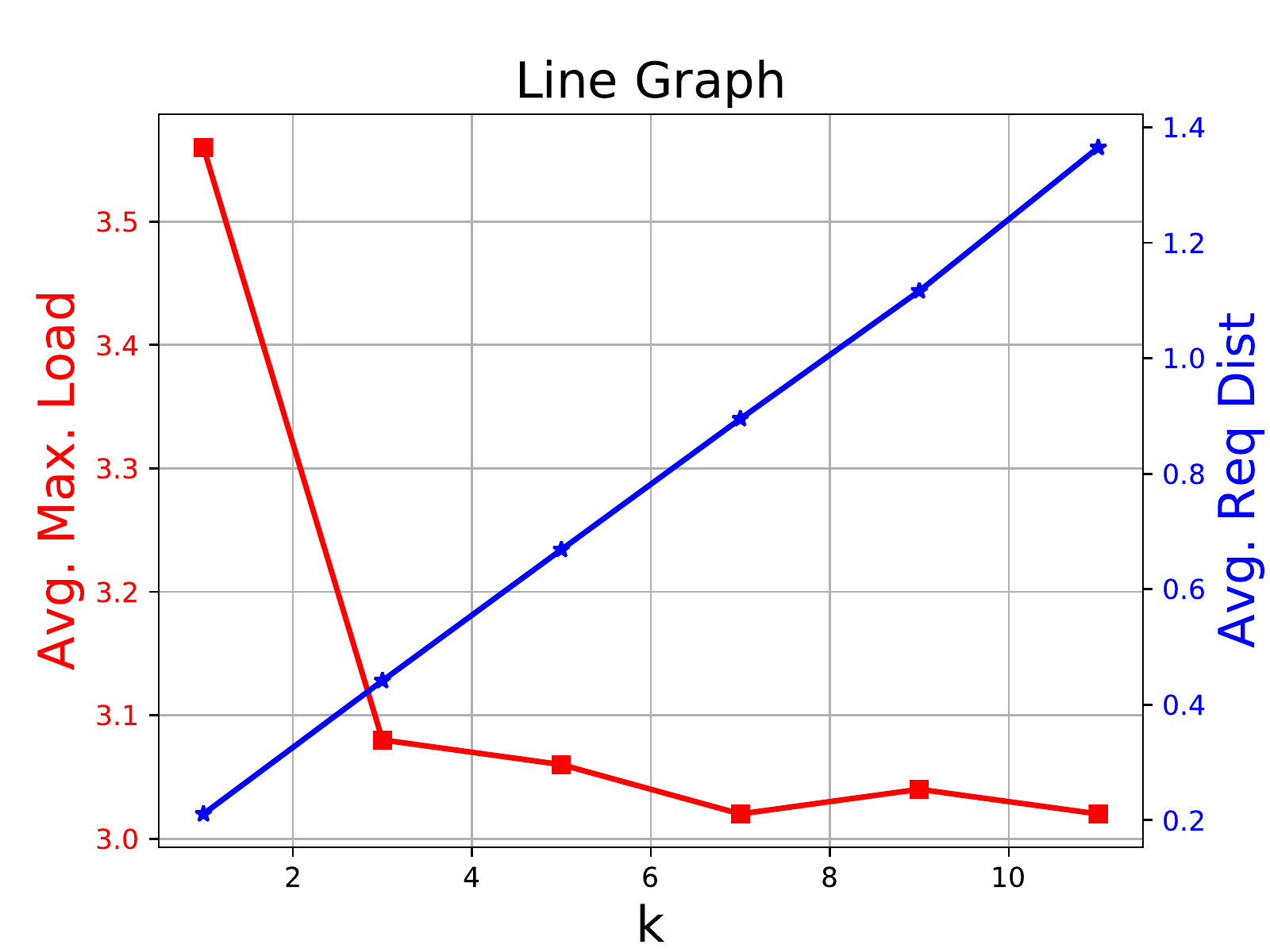}
\vspace{-0.1in}
\caption{Trade-off between average maximum load and average request distance for servers on a Line graph with $m=n=1000$ for Unif-POT($k$) policy under static load balancing system.}
\label{tradeoff}
\vspace{-0.1in}
\end{figure} 
We now define two of the proximity based load balancing policy. Suppose a job arrives at origin server $u\in V.$ Denote $P_u = [p_{uv}, v \in V]$ as the server sampling distribution for the job where $p_{uv}$ is the probability $u$ queries server $v$ for its load information with $p_{uu} = 0.$ We now define the POT policy and two of the proximity based load balancing policies on a graph $G$ as follows. 

\begin{defn}{Power-of-Two (POT) Policy:} 
	If a job arrives at server $u$, then 
	 \beq
	 p_{uv} = 
	 \begin{cases} 
	 \frac{1}{n-1} & \text{if } u\ne v, \\
	 0,       & \text{otherwise }. 
	 \end{cases}
	 \eeq
	That is, a server $v$ is sampled uniformly at random from the remaining $n-1$ servers. The job is then allocated to the server with the smallest load among $u$ and $v.$
	\end{defn}

\begin{defn}{Unif-POT($k$)  Policy:}
	According to this policy, if a job arrives at server $u$, then
	 \beq
	 p_{uv} = 
	 \begin{cases} 
	 \frac{1}{|\mathcal{N}_k(u)|} & \text{if } v \in \mathcal{N}_k(u), \\
	 0,       & \text{otherwise }. 
	 \end{cases}
	 \eeq
That is, a server $v$ is sampled uniformly at random from the set of $k$-hop neighborhood of $u$ out of remaining servers. The job is then allocated to the server with the smallest load among $u$ and $v.$
	\end{defn}

\begin{defn}{InvSq-POT($k$) Policy:}
	According to this policy, if a job arrives at server $u$, then
	\beq
	p_{uv} = 
	\begin{cases} 
		\frac{\Big(\frac{1}{\phi(u,v)^2}\Big)}{\sum\limits_{w \in \mathcal{N}_k(u)}\Big(\frac{1}{\phi(u,w)^2}\Big)} & \text{if } v \in \mathcal{N}_k(u), \\
		0,       & \text{otherwise }. 
	\end{cases}
	\eeq
That is, a server $v \in \mathcal{N}_k(u)$ is sampled  with probability proportional to the inverse square of the distance to $u$. The job is then allocated to the server with the smallest load among $u$ and $v.$
\end{defn}

\begin{remark}\label{re:equivalence}
Observe that Unif-POT($k$) and InvSq-POT($k$) are identical for $k = 1$ when $G$ is undirected and unweighted. Similarly, POT and Unif-POT($k$) are identical for $k = n.$
\end{remark}

\subsection{Maximum Load vs Request Distance Tradeoff}
We first discuss the inherent tradeoff between average maximum load and average request distance for different values of $k$ in Unif-POT($k$) and InvSq-POT($k$)  policies. We perform a simulation experiment with $n=1000$ servers connected through a line graph. We assume $m = 1000$ jobs arrive sequentially to the system and are allocated to servers according to Unif-POT($k$) policy. We report the average of $10$ simulations. We plot both average maximum load and average request distance as a function of neighborhood parameter $k$ as shown in Figure \ref{tradeoff}. We get similar results for the case when allocation is done according to InvSq-POT($k$) policy.

It is clear from Figure \ref{tradeoff} that with increase in the value of $k$ the average maximum load value decreases. This is because the size of $k$-hop neighborhood of an origin server increases as $k$ increases. Thus the load is distributed among a larger group of servers and the behavior of Unif-POT($k$) resembles more and more that of POT policy for large values of $k$. 

However, an increase in $k$ results in increasing values of average request distance. Again, for smaller values of $k,$ the sampled servers remain close to the origin server. However, with increasing values of $k,$ the size of the $k$-hop neighborhood grows. One is more likely to sample a far away server thereby increasing the average request distance. Thus one need to be careful in choosing the correct value of $k$ according to the performance metric one tries to optimize.

\section{Simulation Results: Static Load Balancing Systems}\label{sec:bandb}


\begin{figure*}[htbp]
\hspace{-0.3cm}
\begin{minipage}{0.32\textwidth}
\includegraphics[width=1.1\textwidth]{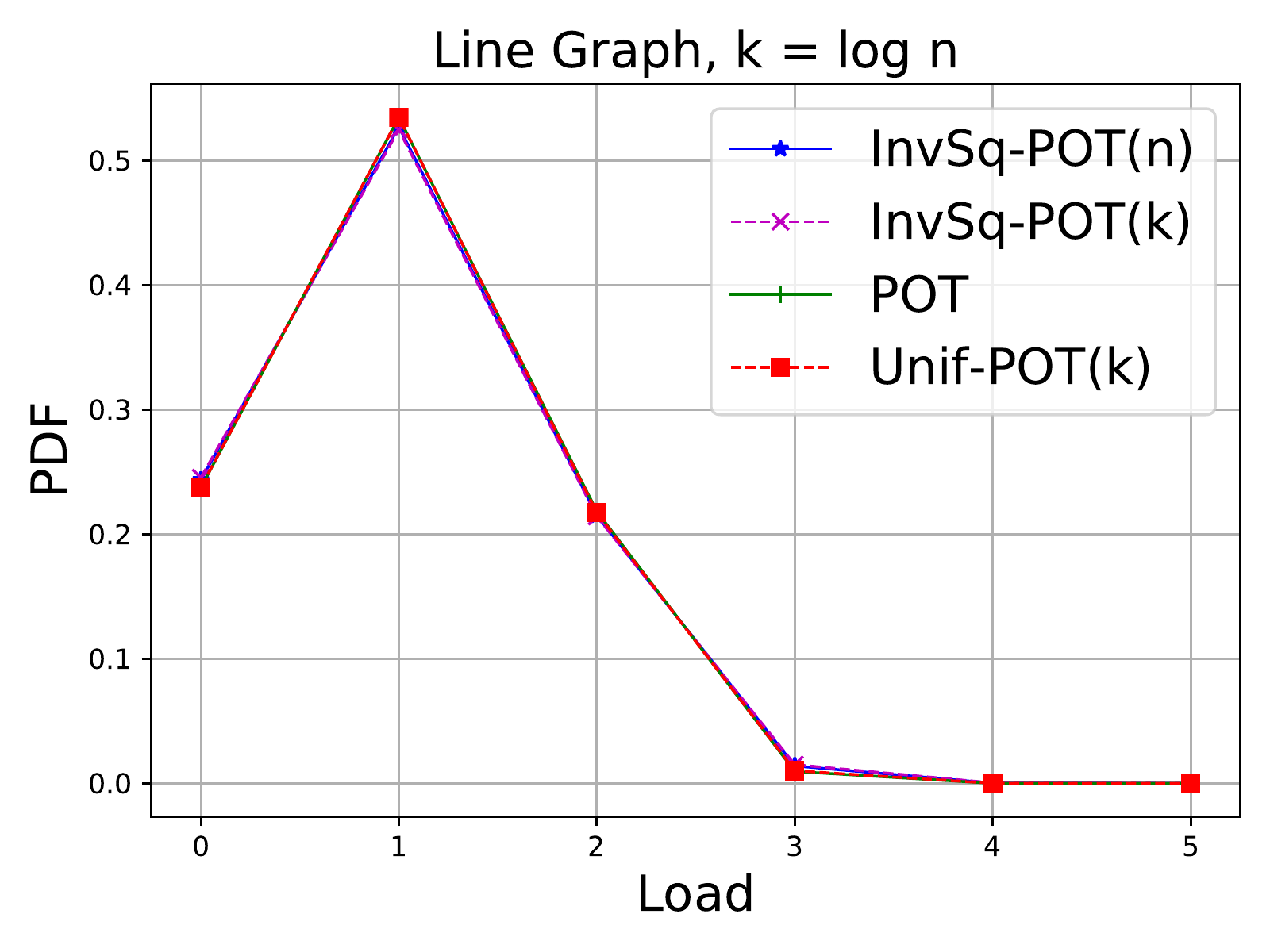}
\subcaption{}
\end{minipage}
\hspace{0.4cm}
\begin{minipage}{0.32\textwidth}
\includegraphics[width=1\textwidth]{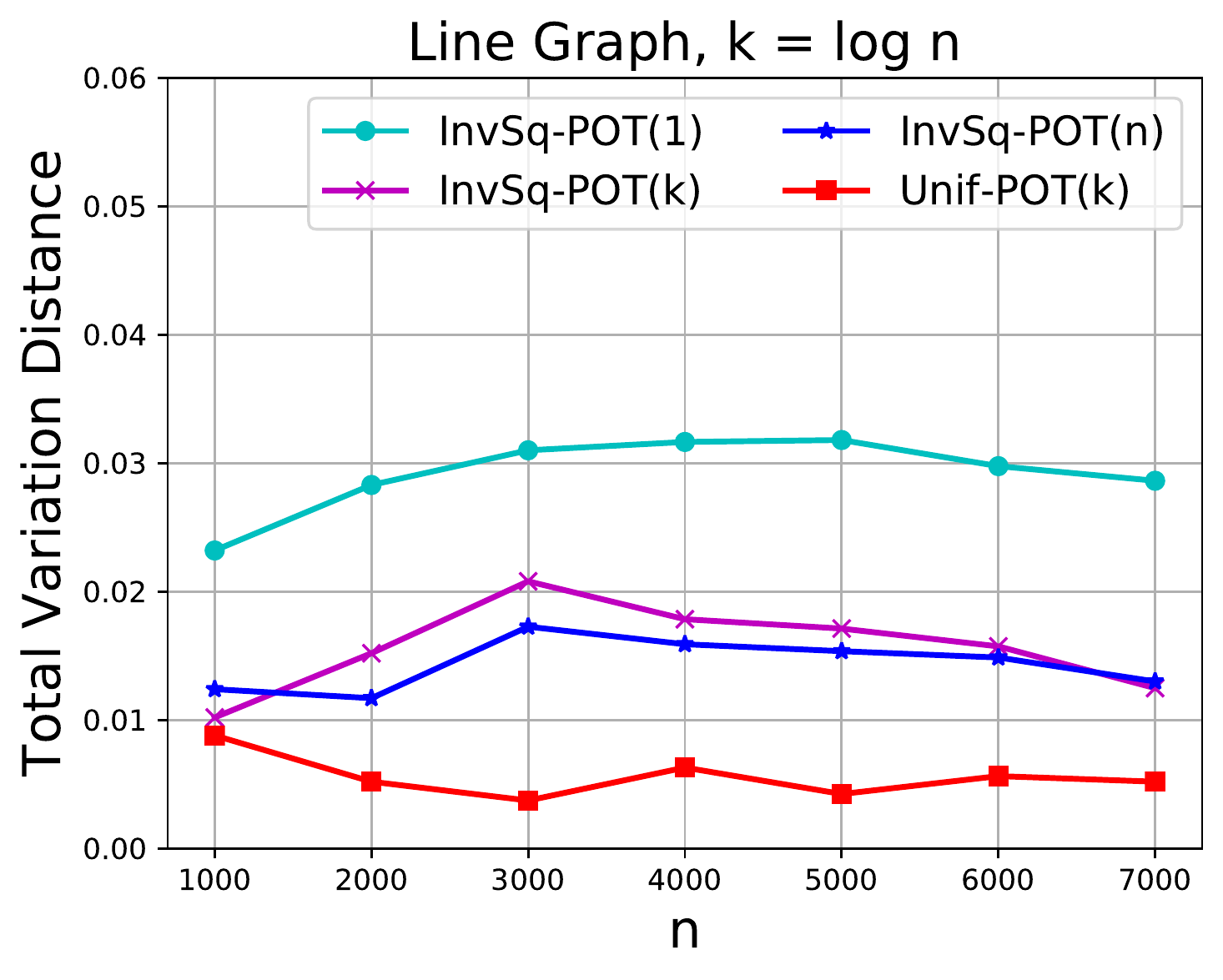}
\subcaption{}
\end{minipage}
\begin{minipage}{0.32\textwidth}
\includegraphics[width=1.14\textwidth]{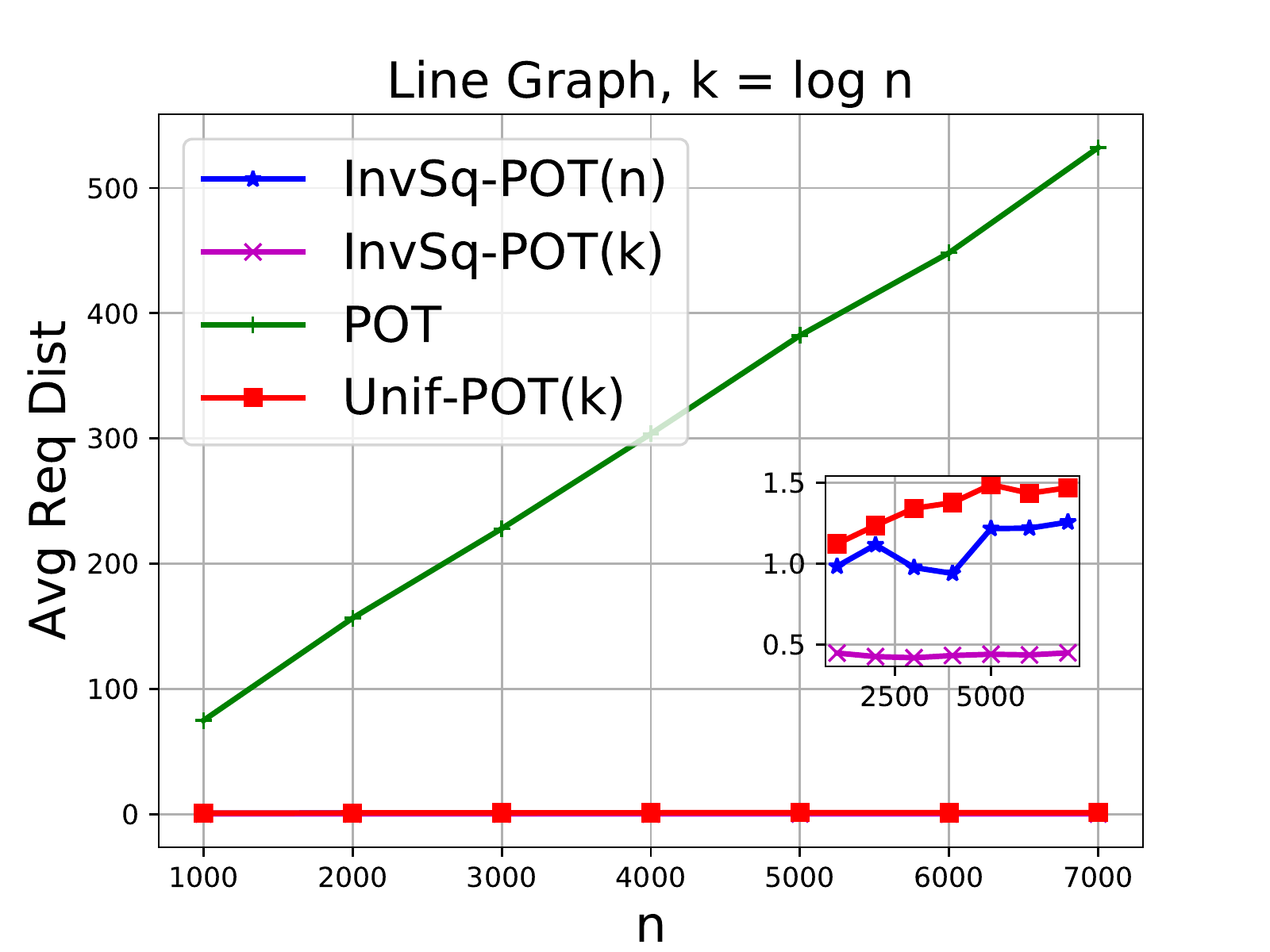}
\subcaption{}
\end{minipage}
\hspace{5cm}
\begin{minipage}{0.32\textwidth}
\includegraphics[width=1.1\textwidth]{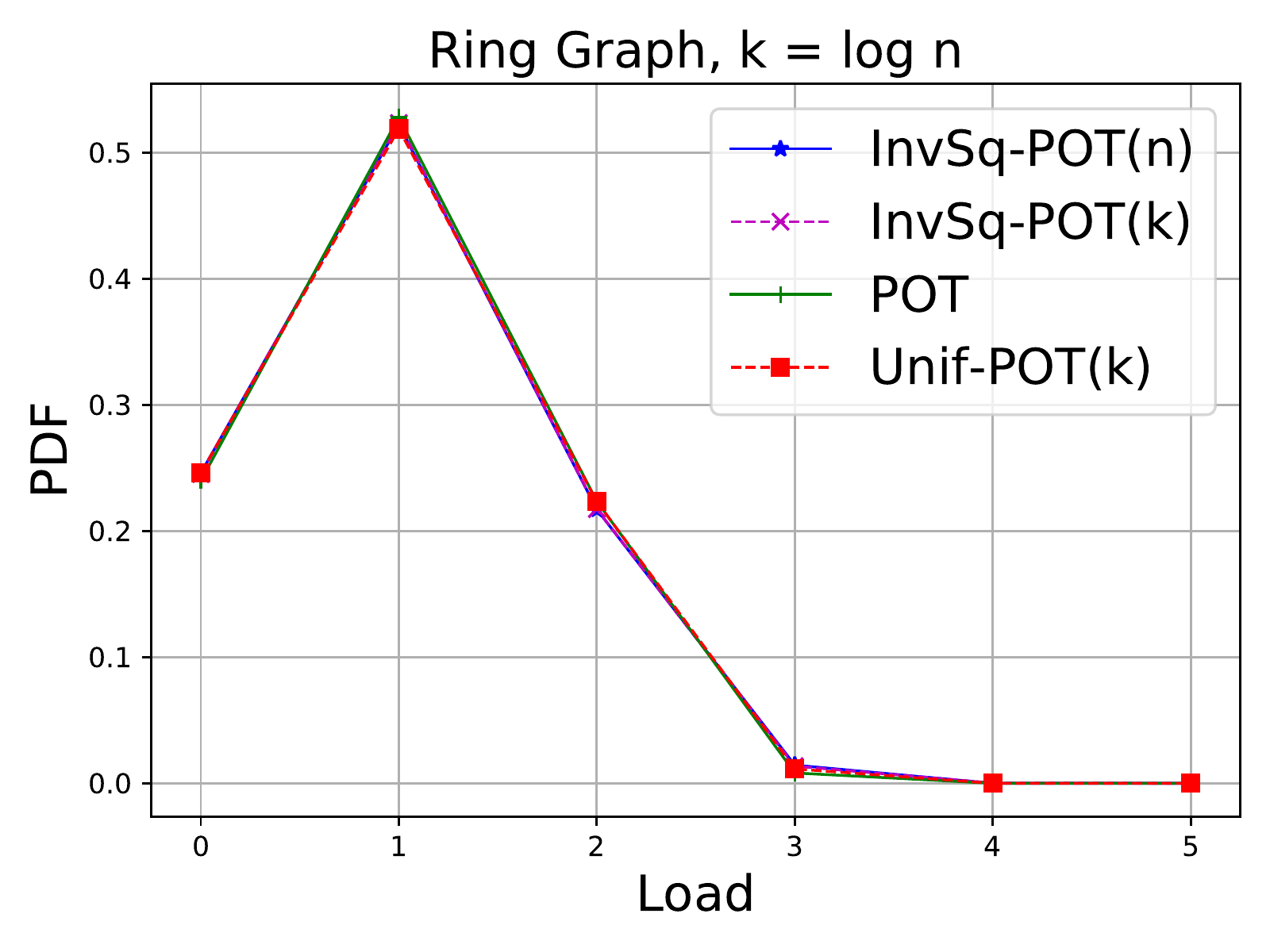}
\subcaption{}
\end{minipage}
\hspace{0.4cm}
\begin{minipage}{0.32\textwidth}
\includegraphics[width=1\textwidth]{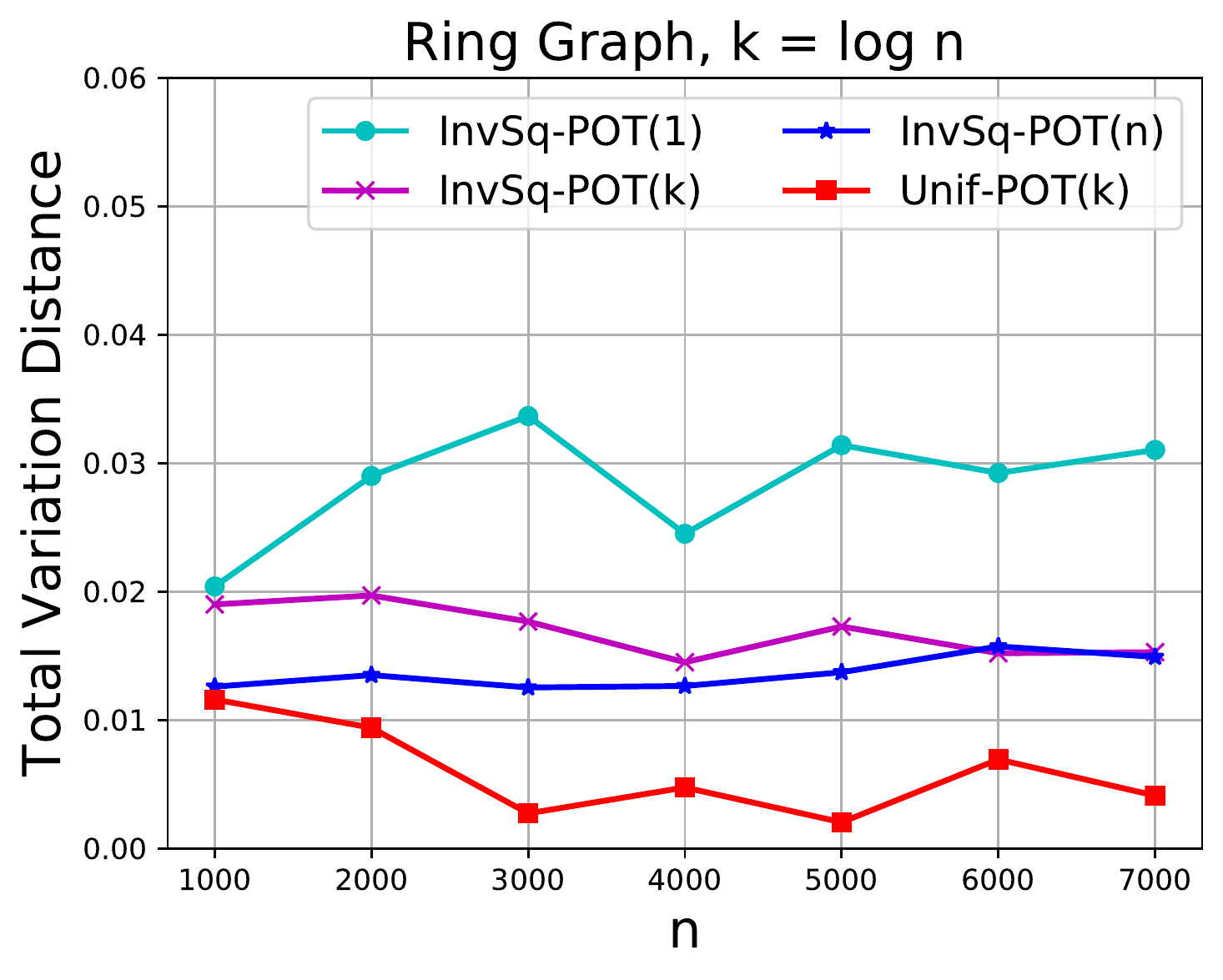}
\subcaption{}
\end{minipage}
\begin{minipage}{0.32\textwidth}
\includegraphics[width=1.14\textwidth]{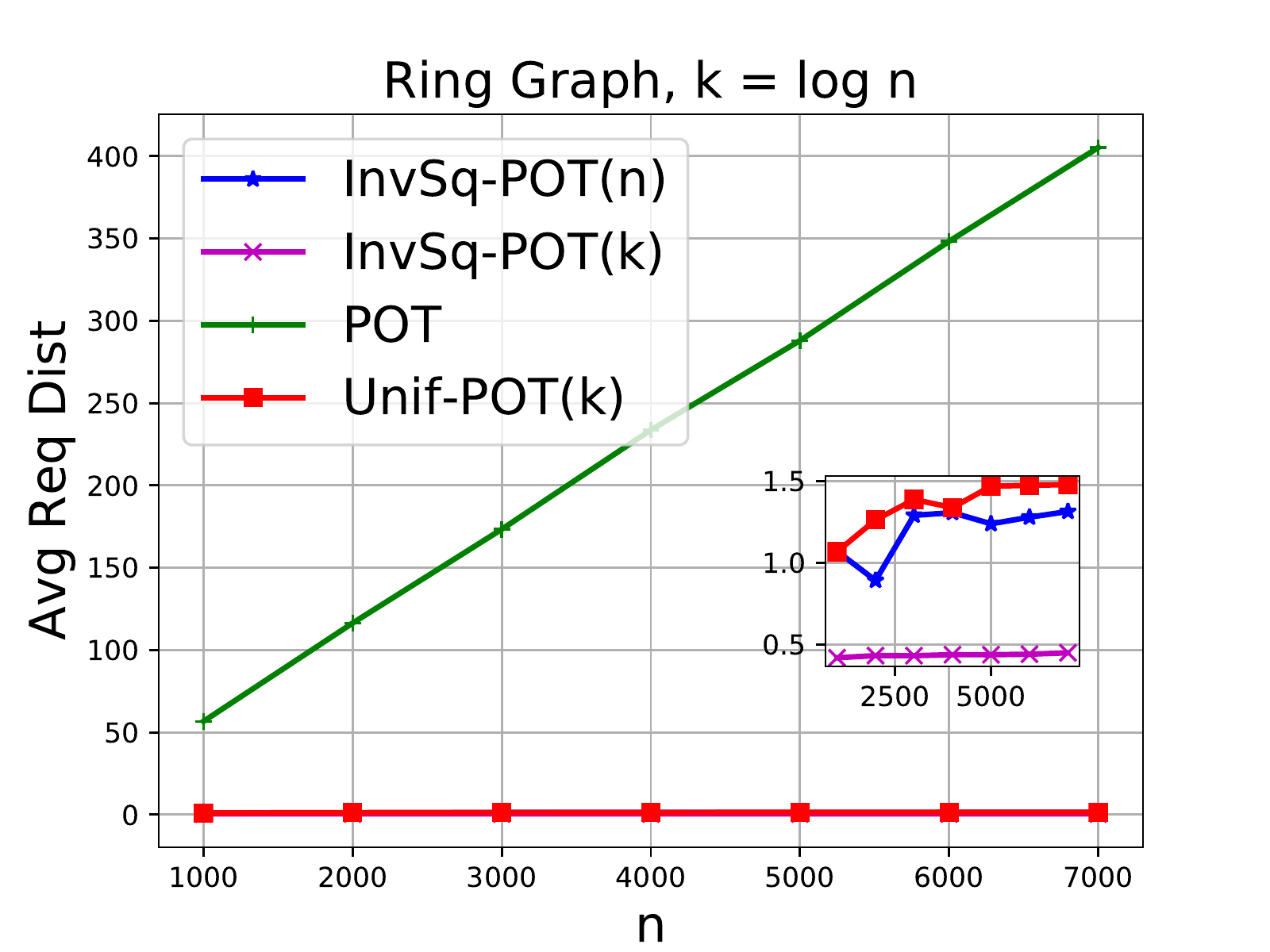}
\subcaption{}
\end{minipage}
\caption{Simulation Results for Unif-POT($k$) and InvSq-POT($k$) for line and ring graphs.}
\label{fig:deter}
\end{figure*}

In this section we present extensive simulation results to illustrate effectiveness of both Unif-POT($k$) and InvSq-POT($k$) policies in static load balancing systems. Our study also provides insights into the choice of a load balancing policy under different load conditions and network topologies.

We implemented the proposed policies using Python programming language to study their performance in a simulated environment. In our study, we evaluated the proposed scheme using both total variation distance to POT load distribution and average request distance as performance metric. To make the performance comparisons between the algorithms meaningful, a number of simulation runs were conducted for each algorithm with different parameter values (e.g., system size, average degree etc.) for underlying graph topology, and the mean metric was selected. We assume the interaction graph remains constant during the simulation. 

If not specified, we assume $n = 10000$ servers interconnected through a graph $G.$ Also, $m = 10000$ jobs arrive to the system sequentially and are allocated to servers under different proximity based POT policies. We set the value of $k = \log n$ in general. We assume the underlying graph $G$ is connected. The results of the simulation experiments are presented in the following sections.

\subsection{Performance Comparison for Deterministic Graphs}
In this Section, we analyze the performance of fixed degree deterministic graphs: Line and Ring. The results are presented in Figure \ref{fig:deter}. First we plot the PDF of load for Line and Ring topologies in Figure \ref{fig:deter} (a) and (d). We compare the PDFs for POT,  Unif-POT($\log n$),  InvSq-POT($\log n$)  and InvSq-POT($n$) policies. Surprisingly, the PDFs of Unif-POT($\log n$),  InvSq-POT($\log n$)  and InvSq-POT($n$) almost exactly match the PDF of POT for both Line and Ring topologies. Also note that, the maximum load only occurs at a very small fraction of servers. Thus it is important, in practice, to consider policies that have more mass in the middle of load distribution curve than the ones that have the lowest maximum load.

Next, we compare the proximity based policies to POT with respect to total variation distance for both Line and Ring topologies. We plot total variation distance between load distributions of proximity based policies and POT as a function of number of servers in Figures \ref{fig:deter} (b) and (e). Again to our surprise, all proximity based policies with $k = \log n, n$ achieve total variation distances as low as $2\%$ across a wide range of values of $n.$ Also note that, InvSq-POT($1$) achieves a load distribution farthest from POT while Unif-POT($\log n$) achieves the closest. Due to its uniform way of sampling, Unif-POT($\log n$) achieves the lowest total variation distance. However, due to bias towards closest severs, both InvSq-POT($\log n$)  and InvSq-POT($n$) achieve higher variation distances.  Due to load-communication cost trade-off, a very local policy InvSq-POT($1$), achieves even higher variation distance. Both InvSq-POT($\log n$) and InvSq-POT($n$) seem to converge to a particular variation distance for both Line and Ring topologies for high values of $n$.

Finally we plot average request distance as a function of number of servers as shown in Figures \ref{fig:deter} (c) and (f). The average path length can be thought of as an upper bound and in positive correlation with average request distance under POT policy, i.e. higher values of average path length imply higher values for average request distance under POT policy. Surprisingly, proximity based policies significantly decrease  average request distance ($\sim99\%$ reduction) for high values of $n$.   Since average path length for both line and ring graphs scale as $O(n),$ the average request distance also drastically increases with increase in $n.$ Also note that, InvSq-POT($\log n$) achieves the lowest average request distance compared to other policies.

\subsection{Performance Comparison for Random Graphs}
\begin{figure*}[!htbp]
\begin{minipage}{0.32\textwidth}
\includegraphics[width=1\textwidth, height=0.8\textwidth]{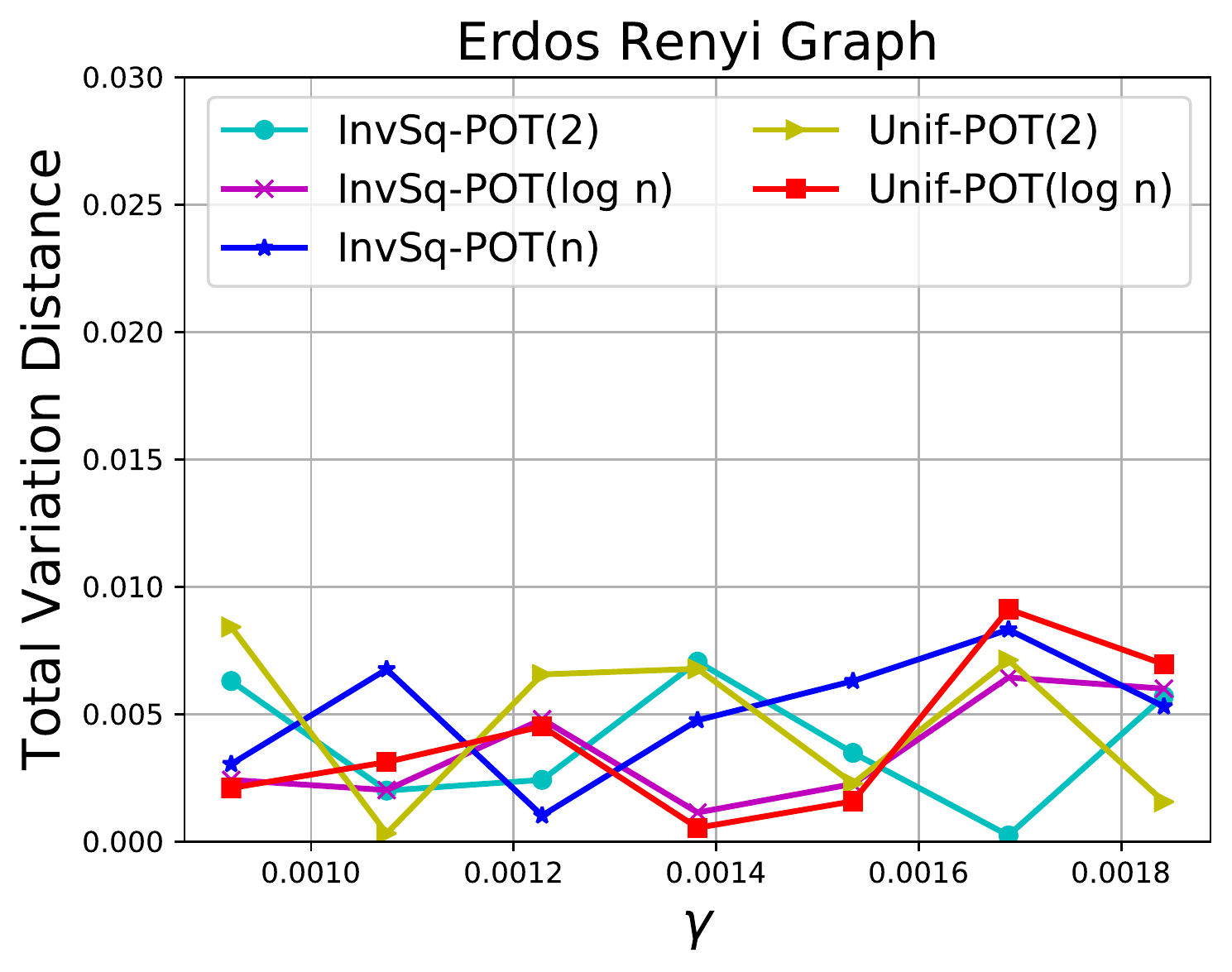}
\subcaption{}
\end{minipage}
\begin{minipage}{0.32\textwidth}
\includegraphics[width=1.1\textwidth]{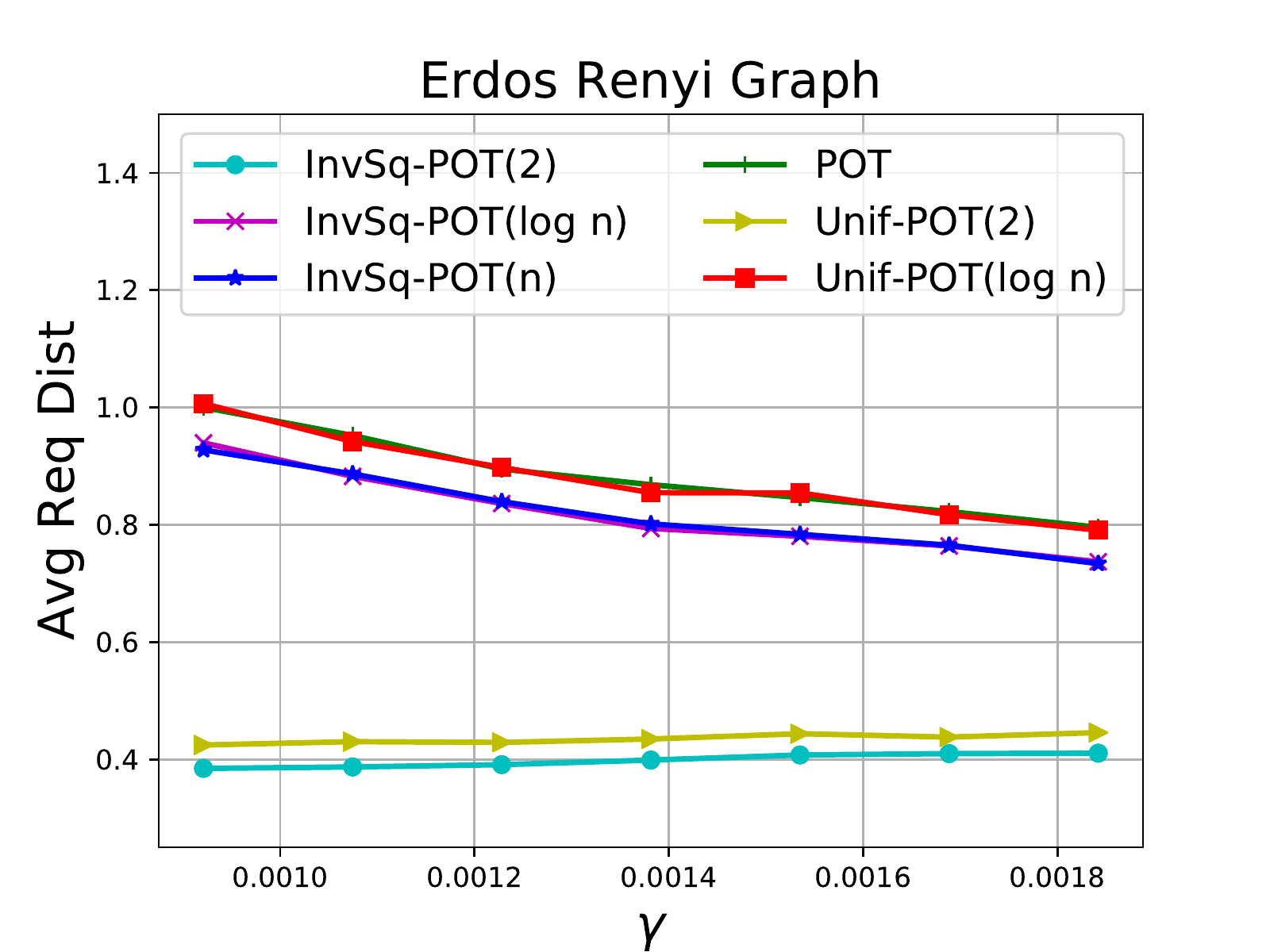}
\subcaption{}
\end{minipage}
\begin{minipage}{0.32\textwidth}
\includegraphics[width=1.1\textwidth]{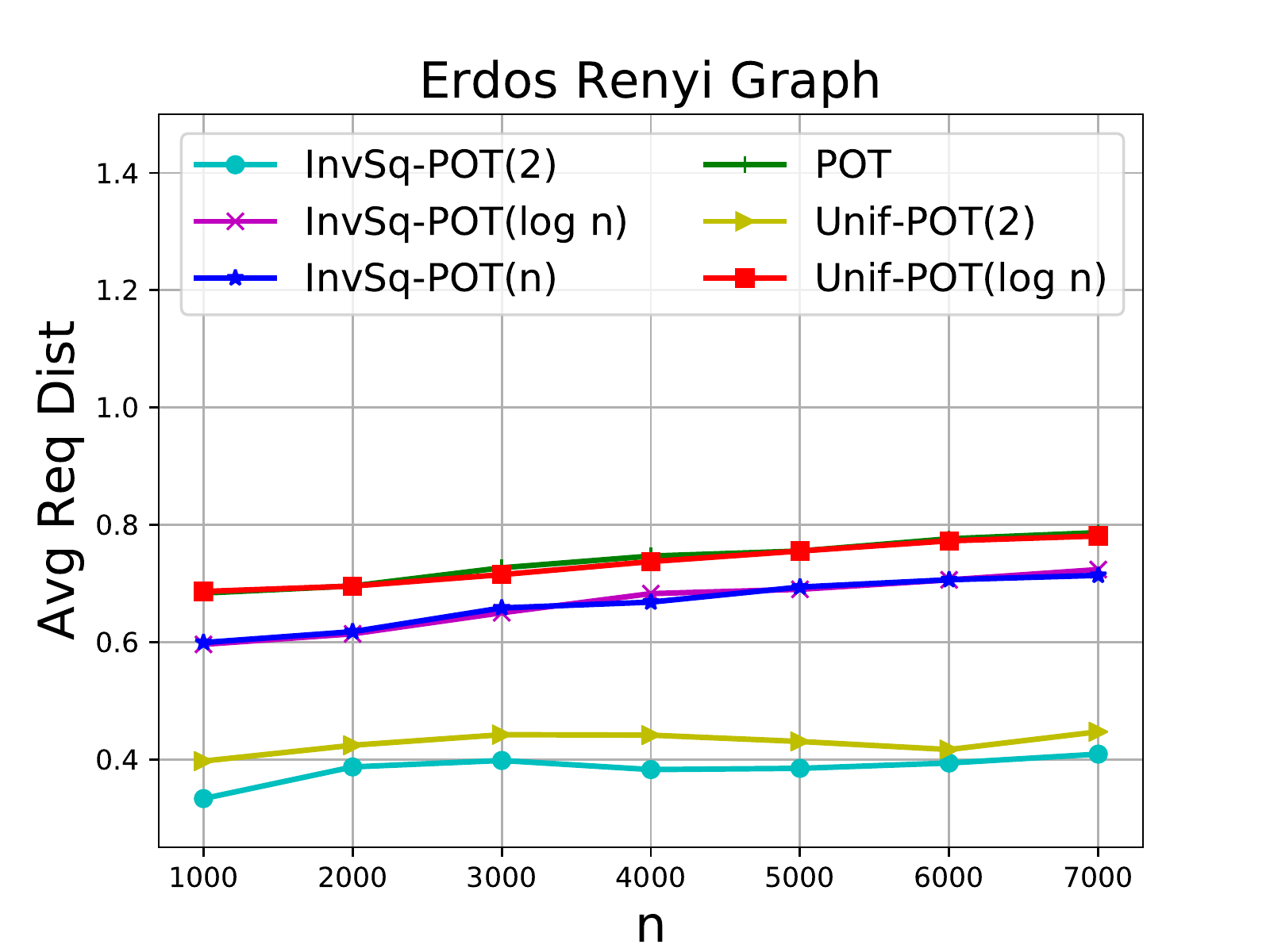}
\subcaption{}
\end{minipage}
\begin{minipage}{0.32\textwidth}
\includegraphics[width=1\textwidth]{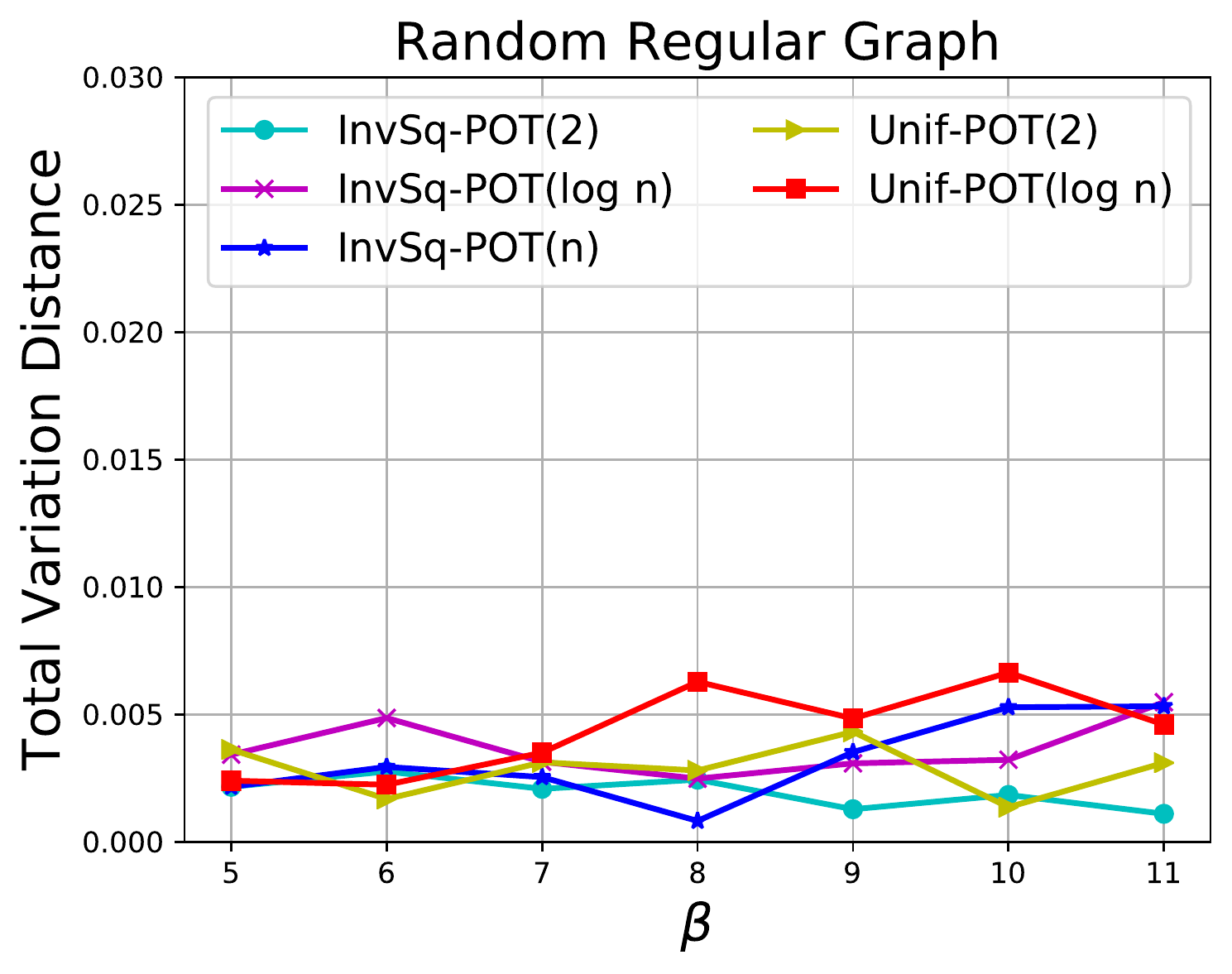}
\subcaption{}
\end{minipage}
\begin{minipage}{0.32\textwidth}
\includegraphics[width=1.1\textwidth]{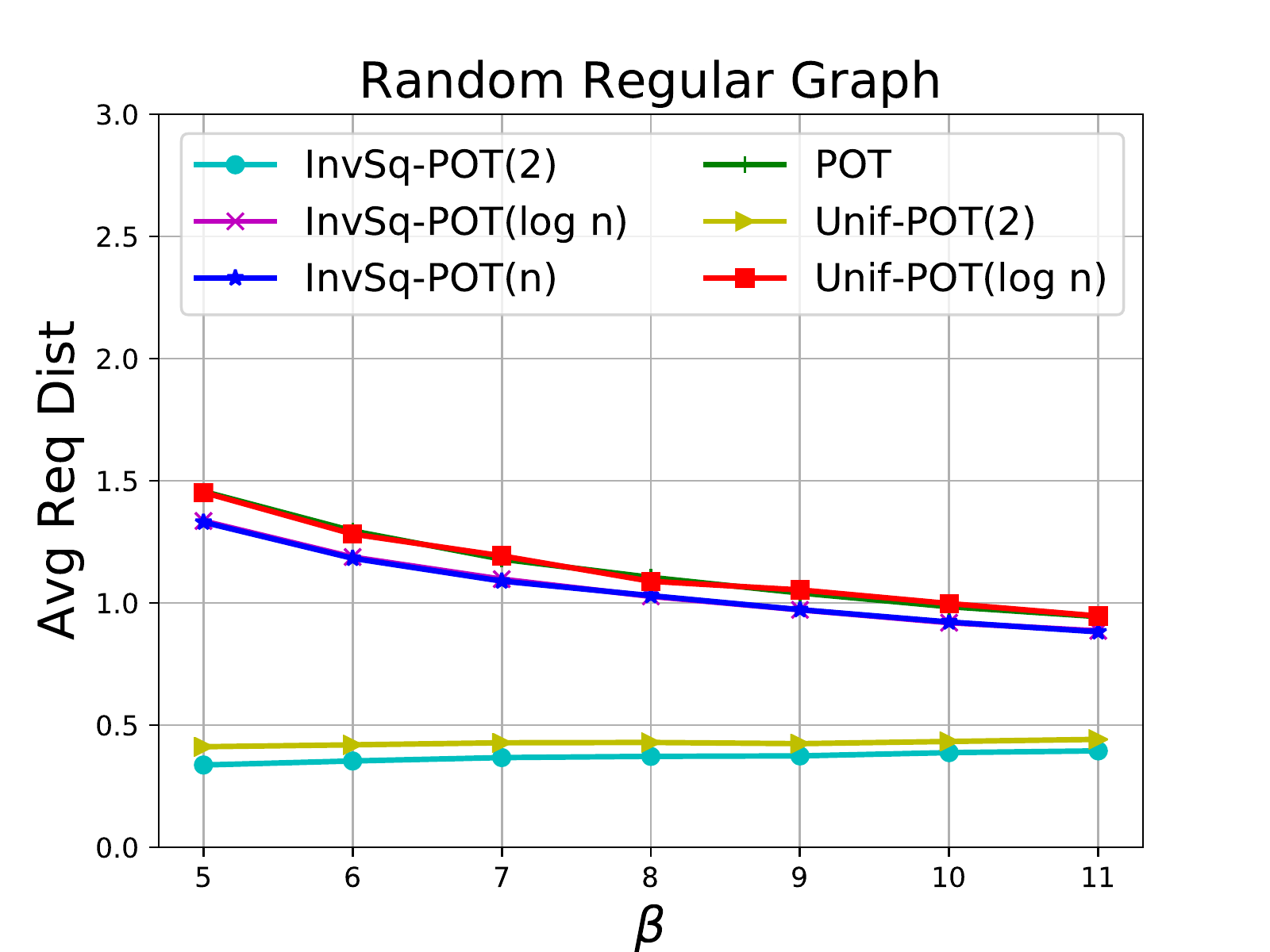}
\subcaption{}
\end{minipage}
\begin{minipage}{0.32\textwidth}
\includegraphics[width=1.1\textwidth]{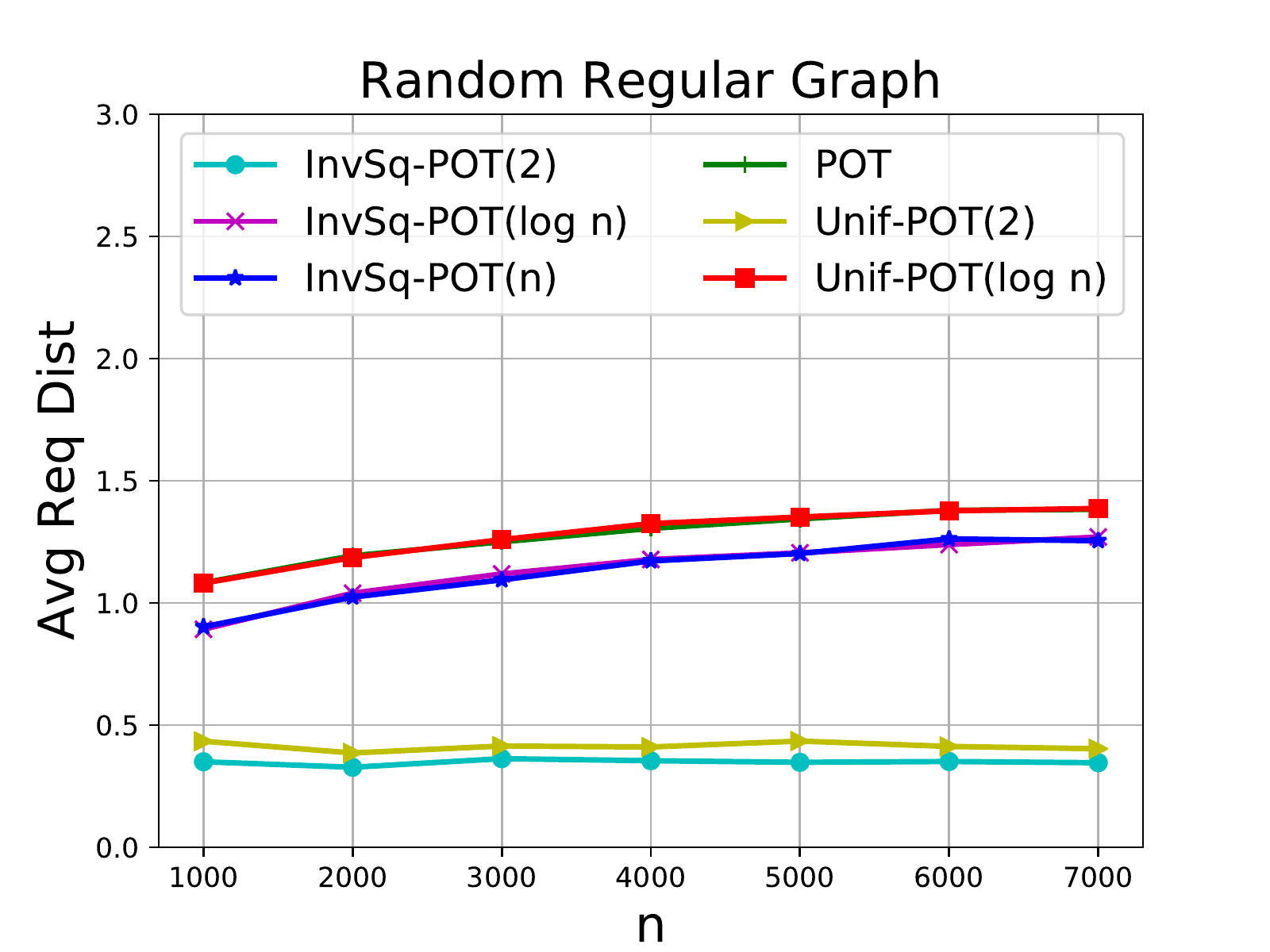}
\subcaption{}
\end{minipage}
\begin{minipage}{0.32\textwidth}
\includegraphics[width=1\textwidth]{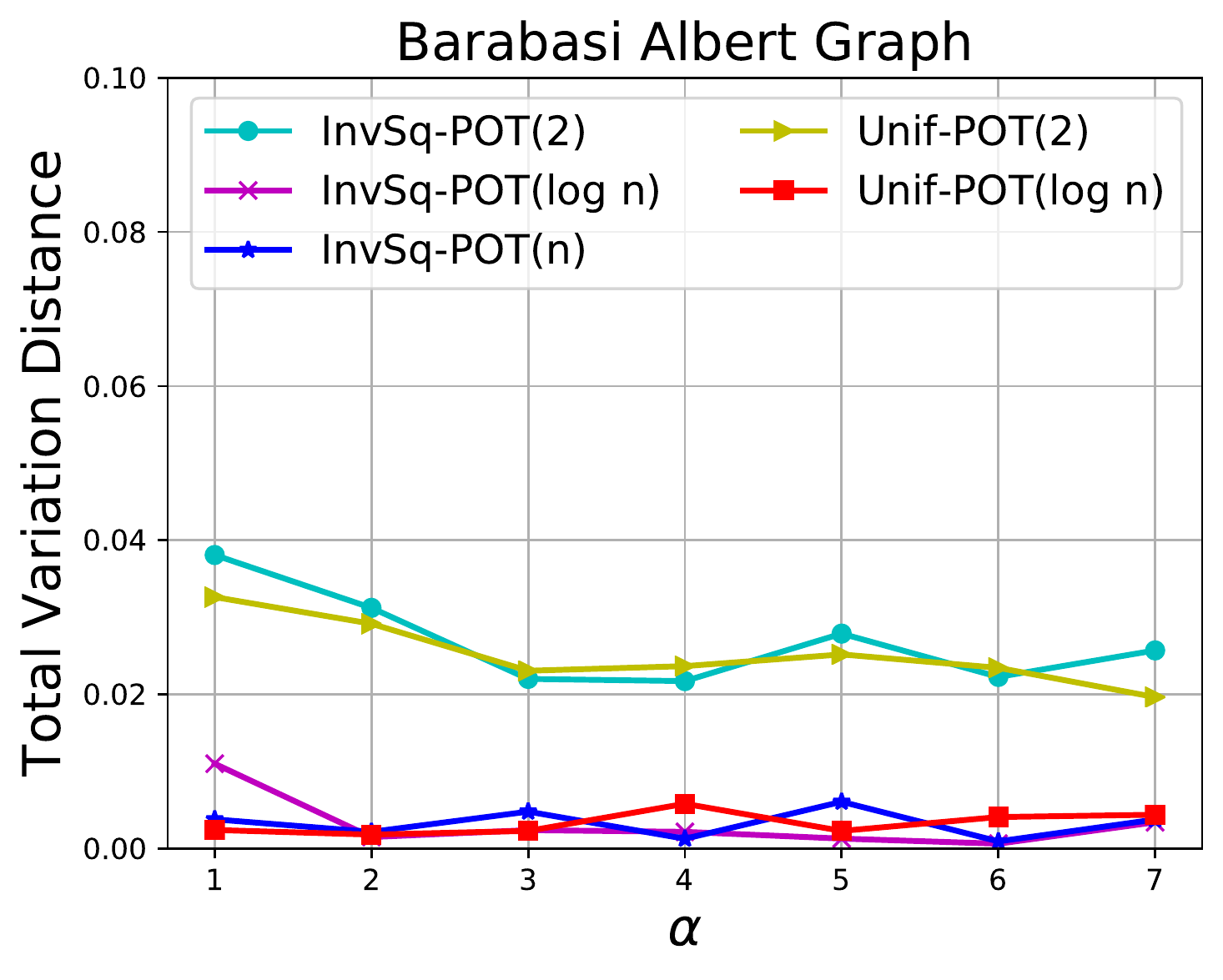}
\subcaption{}
\end{minipage}
\begin{minipage}{0.32\textwidth}
\includegraphics[width=1.1\textwidth]{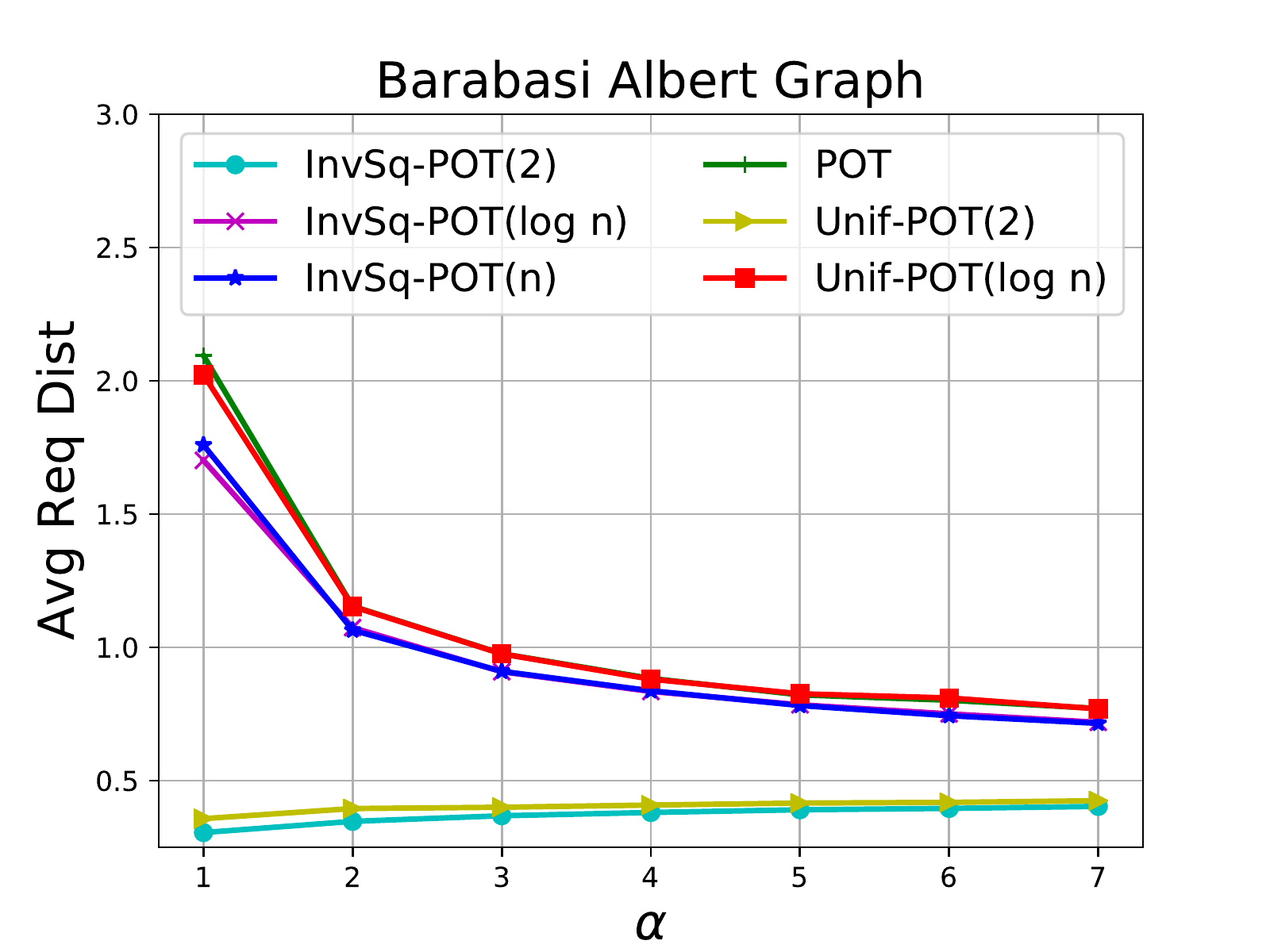}
\subcaption{}
\end{minipage}
\begin{minipage}{0.32\textwidth}
\includegraphics[width=1.1\textwidth]{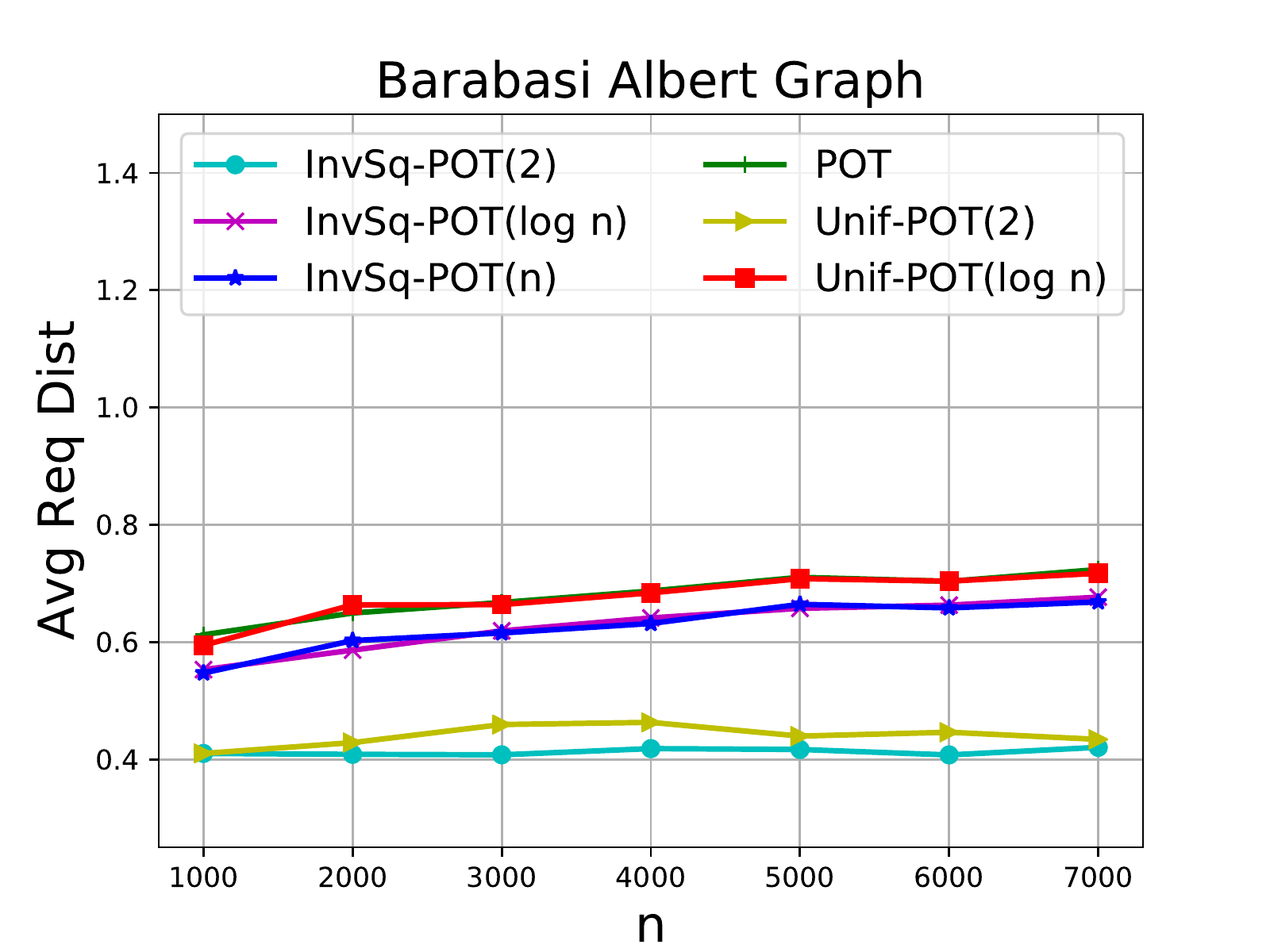}
\subcaption{}
\end{minipage}
\vspace{-0.1in}
\caption{Simulation Results for Unif-POT($k$) and InvSq-POT($k$) with $n=10000$ and $k = 2,\log n,n$ for random graphs.}
\label{random}
\end{figure*}

\begin{figure*}[!htbp]
\centering
\hspace{-0.3cm}
\begin{minipage}{0.32\textwidth}
\includegraphics[width=1\textwidth]{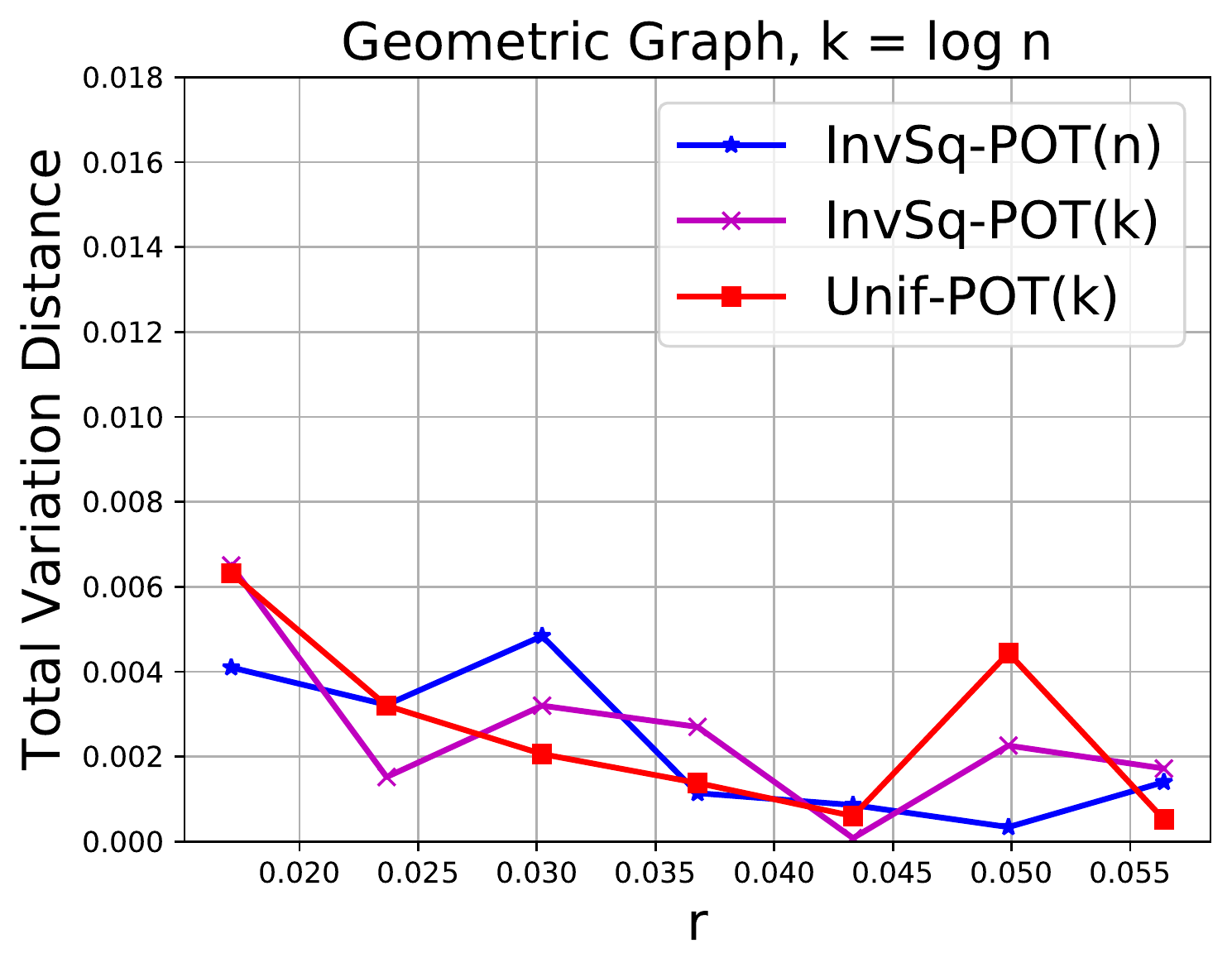}
\subcaption{}
\end{minipage}
\begin{minipage}{0.32\textwidth}
\includegraphics[width=1.1\textwidth]{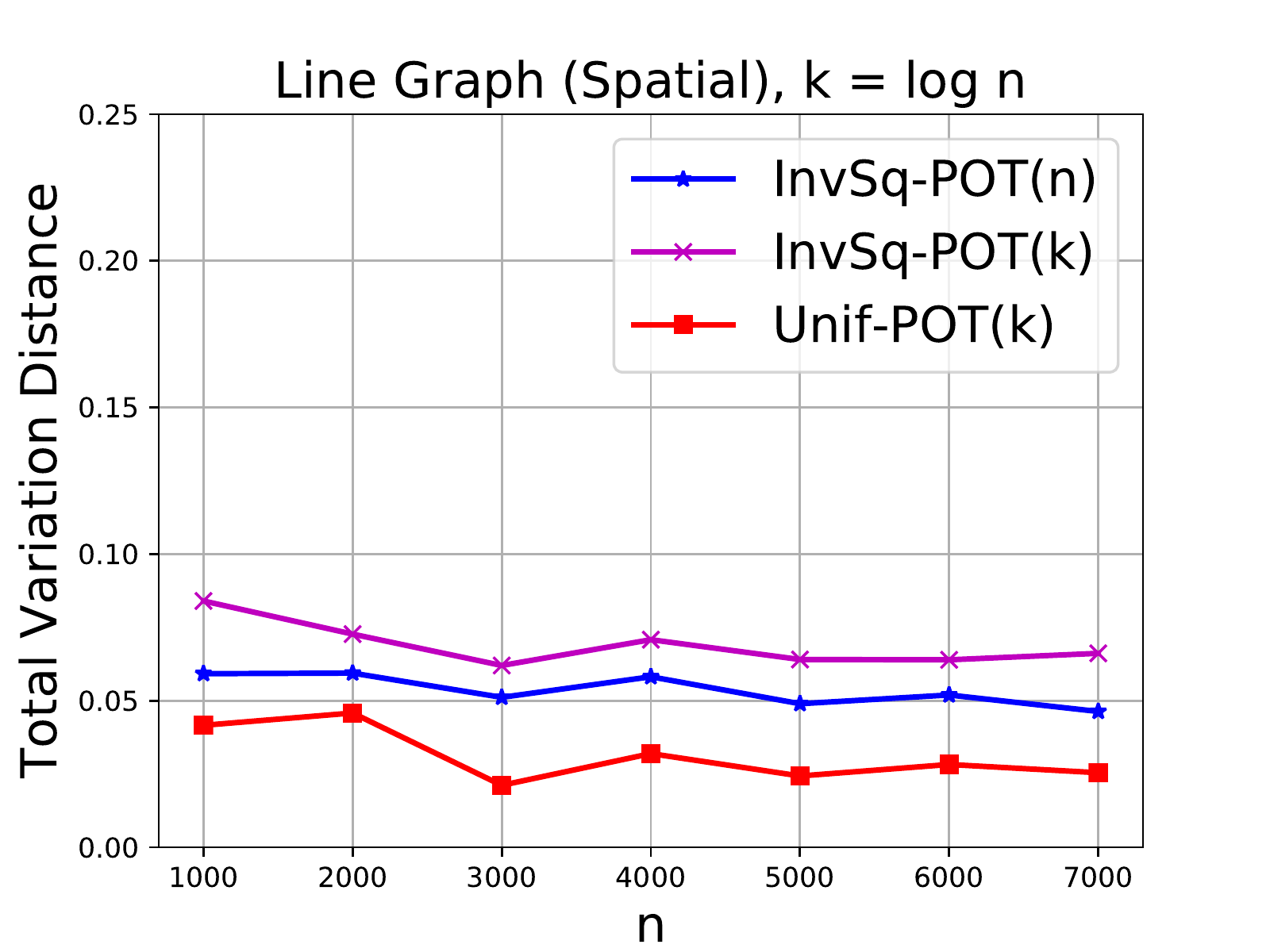}
\subcaption{}
\end{minipage}
\begin{minipage}{0.32\textwidth}
\includegraphics[width=1.1\textwidth]{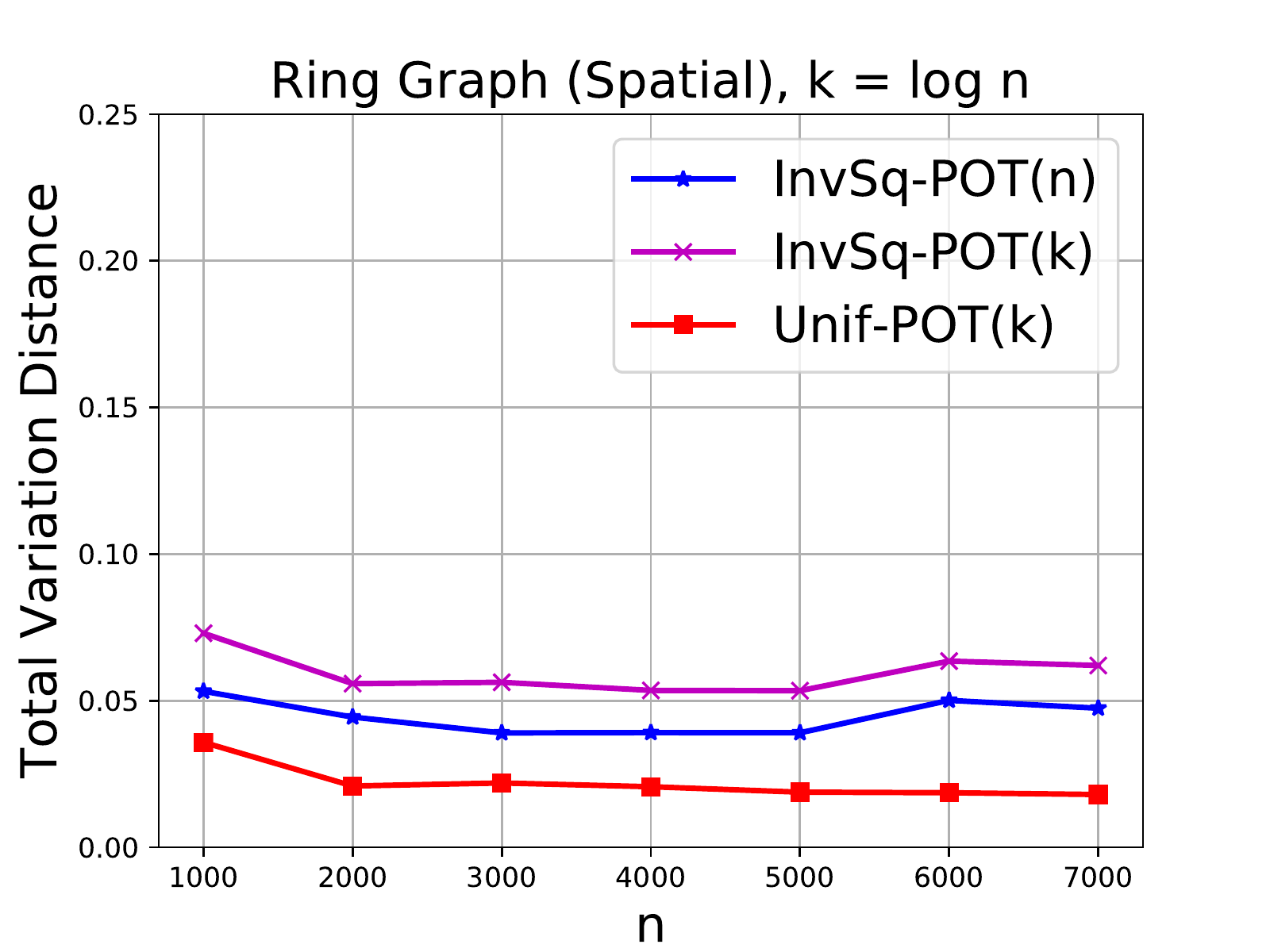}
\subcaption{}
\end{minipage}
\begin{minipage}{0.32\textwidth}
\includegraphics[width=1.1\textwidth]{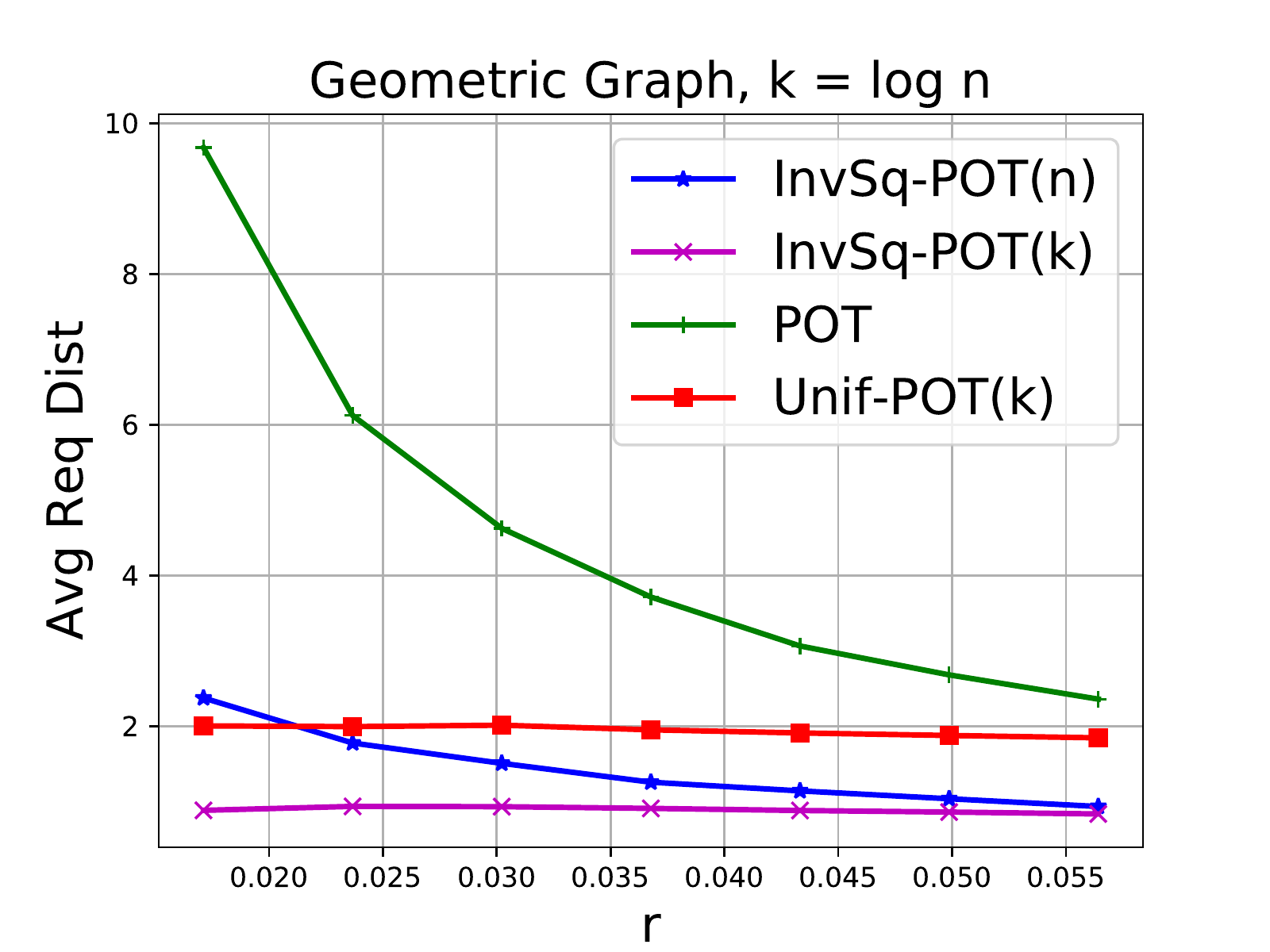}
\subcaption{}
\end{minipage}
\begin{minipage}{0.32\textwidth}
\includegraphics[width=1.1\textwidth]{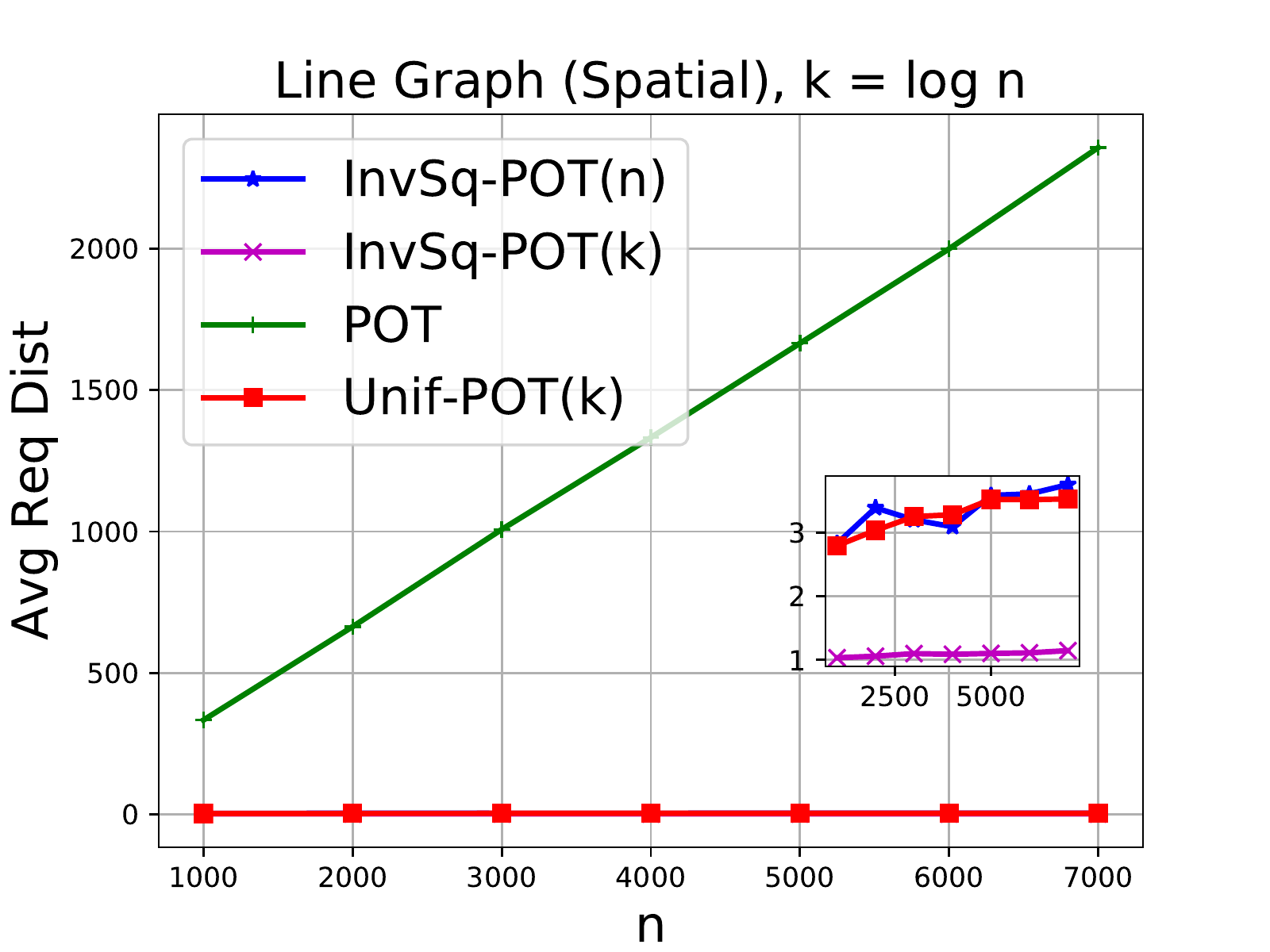}
\subcaption{}
\end{minipage}
\vspace{-0.1in}
\caption{Simulation Results for Unif-POT($k$) and InvSq-POT($k$) with $n=10000$ and $k = \log n,n$ for spatial graphs.}
\label{spatial}
\end{figure*}

\begin{table}
	\begin{center}
		\begin{tabular}{ |c|c|c| c|} 
			\hline
			Graph Type &  $n$ & $m$ & Parameters\\ 
			\hline
			Erdos Reny $(n, \gamma)$ & $10000$& $10000$& $\gamma: [\log n/n,\cdots, 2\log n/n]$\\ 
			Random Regular $(n, \beta)$ & $10000$& $10000$&$\beta: [5, 6,\cdots,11]$ \\
			Barabasi Albert $(n, \alpha)$ & $10000$&$10000$&$\alpha: [1,2,\cdots,7]$\\
			\hline
		\end{tabular}
		\caption{Simulation parameters for random graph topologies.}
		\label{table:smrnd}
	\end{center}
\end{table}

We now move our focus to studying the behavior of the proximity based policies on random graphs. In particular, we compare the performance of Unif-POT($k$) and InvSq-POT($k$) policies for $k = 2,\log n, n$ to that of POT policy. We consider the three random graphs considered in the simulation: Erdos Reny (ER), Random Regular (RR) and Barabasi Albert (BA) are connected. We present the system and network parameters used in the simulation in Table \ref{table:smrnd}. The results are presented in Figure \ref{random}.

We first plot the total variation distance of proximity based policies as a function of ER edge probability parameter $\gamma$ as shown in Figure \ref{random} (a). Note that, for all values of $k = 2,\log n, n$, both the proximity based policies produce a variation distance as low as $0.5\%.$ This is surprising since with $k = O(1) = 2$ we only sample two hop neighborhood of the origin server, i.e. InvSq-POT($2$) is a very local policy. But we are able to produce load distribution behavior almost identical to that of POT policy which samples from the entire set of servers. We observe similar trend for RR graph as well, as shown in Figure \ref{random} (d). But, we observe different results for the BA graph as shown in Figure \ref{random} (g). For BA graph, we observe that $k = 2$ for both proximity based policies produce higher total variation distance as compared to $k=\log n, n.$ However, the variation distance is still small for $k = 2$ for both policies fluctuating around $3\%.$ Note that, an increase in the value of $k$ ideally should decrease variation distance since the sampling set size increases with $k.$ Observe that the variation distance of policies under a Line or Ring topology is higher than that of any random topologies with fixed degree (Ex: RR topology). Higher graph densities in random topologies yield lower variation distance compared to a Line or Ring topology.

We now study the effect of the network parameter on the average request distances of the proximity based policies as shown in Figures \ref{random} (b), (e), (h). First observe that an increase in the value of network parameters ($\alpha, \beta$ and $\gamma$) increases the graph density of the corresponding graphs (BA, RR and ER) and hence connectedness. This results in decrease in average request distances. Also observe the insensitivity of proximity based policies with $k=2$ to the network size $n$. As expected, proximity policies with $k=2$ produce very low request distances as compare to the case when $k = \log n.$

Next, we study the scalability of average request distance with respect to network size as shown in Figures \ref{random} (c), (f) and (i). Note that the average path length ER and BA exhibits small ($\log n$) and ultra small world ($\log n/\log \log n$) behavior respectively \cite{Fronczak2004}. We observe similar logarithmic growth for the average request distance for POT policy as evident from Figures \ref{random} (c) and (i). Due to small world behavior, the observed average request distances are pretty small for a BA or ER topology across all policies as compared to Line and Ring topologies of similar network size. Again as expected, proximity policies with $k = 2$ are insensitive to changes in network size and produce the smallest average request distances. Also, observe that between Unif-POT($k$) and InvSq-POT($k$) for every $k,$  InvSq-POT($k$) policies produce lower average request distances for networks of similar size.

\subsection{Performance Comparison for Spatial Graphs}

\begin{table}
	\begin{center}
		\begin{tabular}{ |c|c|c| c|} 
			\hline
			Graph Type &  $n$ & $m$ & Parameters\\ 
			\hline
			Random Geometric $(n, r)$ & $10000$& $10000$& $r: [\sqrt{\log n/\pi n},\cdots, \sqrt{\sqrt{n}/\pi n}]$\\ 
			Spatial Line $(n, L_{max})$ & $[1000,\cdots,7000]$& $[1000,\cdots,7000]$&$L_{max}: [1000,\cdots,7000]$ \\
			Spatial Ring $(n, R)$ & $[1000,\cdots,7000]$&$[1000,\cdots,7000]$&$R: 1$\\
			\hline
		\end{tabular}
		\caption{Simulation parameters for spatial graph topologies.}
		\label{table:smspatial}
	\end{center}
\end{table}

\begin{figure*}[!htbp]
\centering
\hspace{-0.3cm}
\begin{minipage}{0.32\textwidth}
\includegraphics[width=1\textwidth]{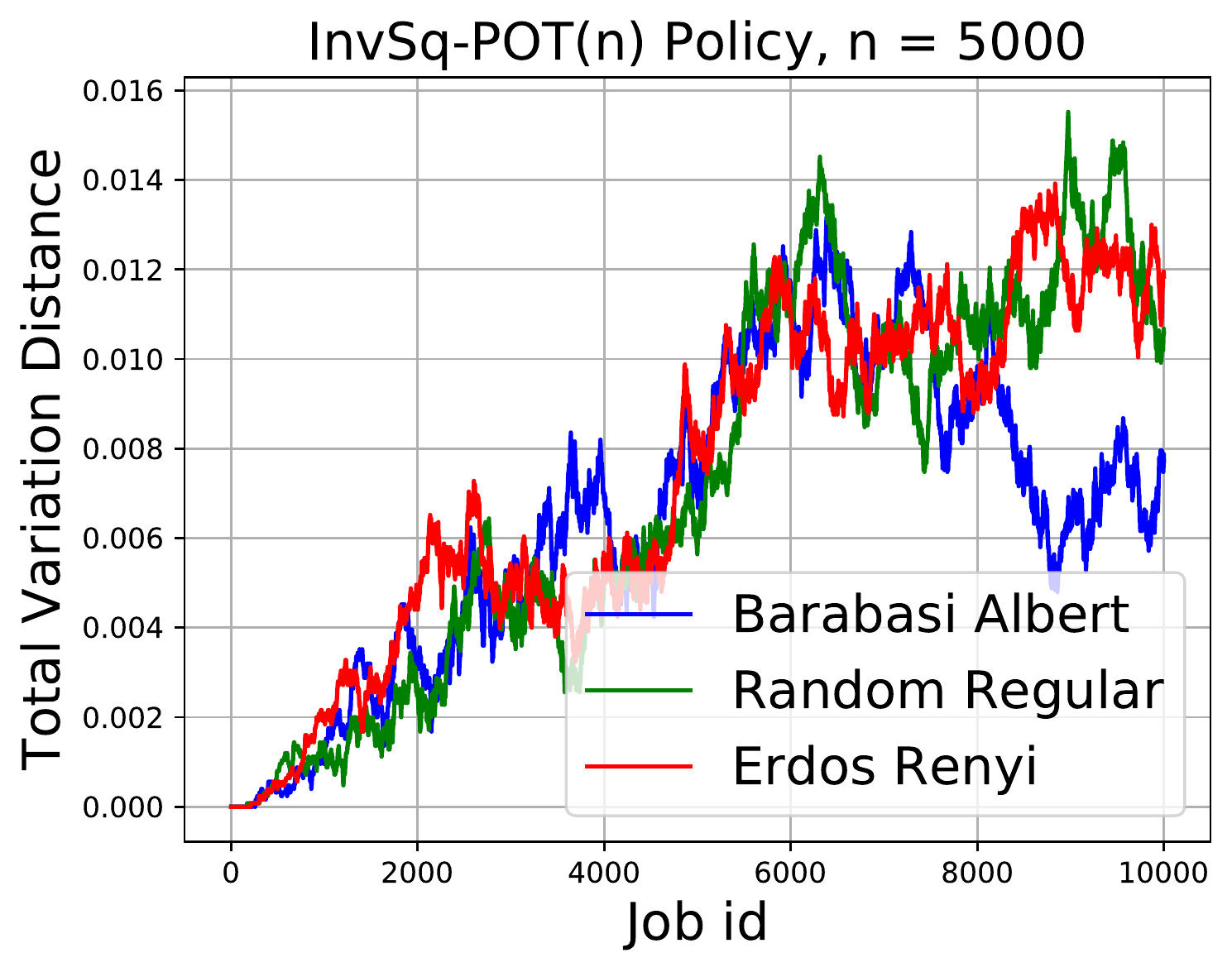}
\subcaption{}
\end{minipage}
\begin{minipage}{0.32\textwidth}
\includegraphics[width=1\textwidth]{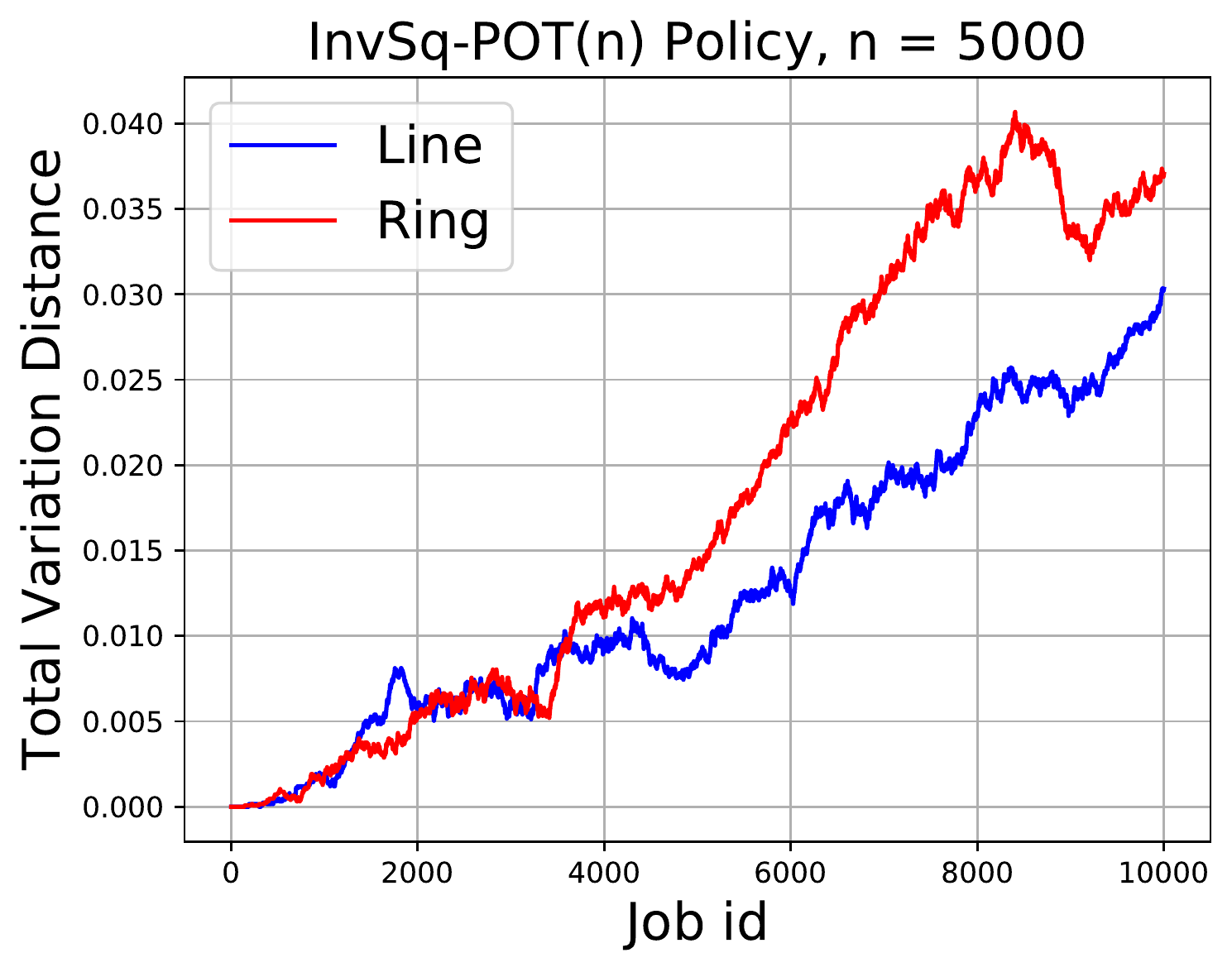}
\subcaption{}
\end{minipage}
\begin{minipage}{0.32\textwidth}
\includegraphics[width=1.1\textwidth]{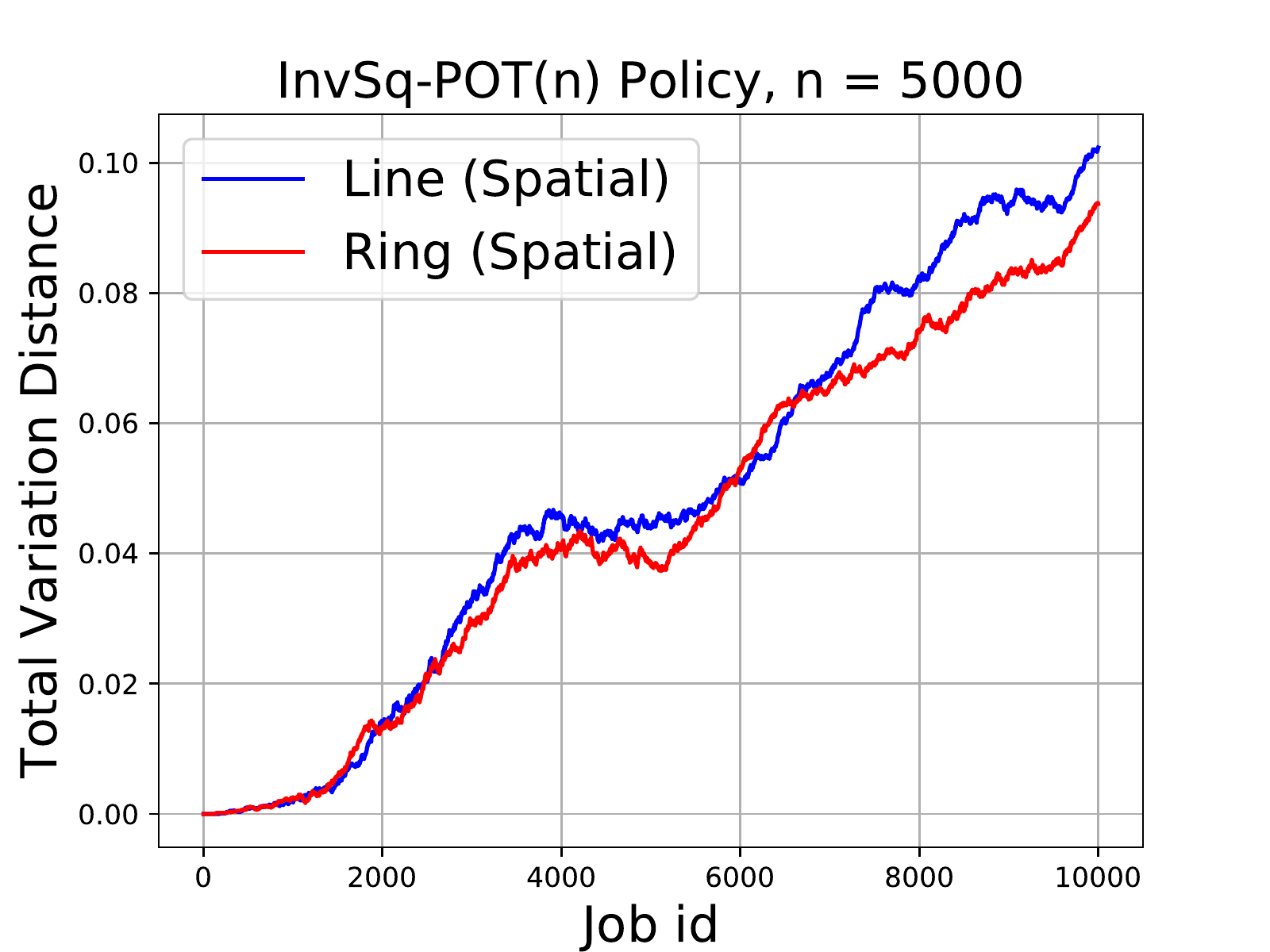}
\subcaption{}
\end{minipage}
\vspace{-0.1in}
\caption{Simulation Results for evolution of total variation distance for InvSq-POT($n$) policy.}
\label{evolution}
\end{figure*}

\begin{figure}[!htbp]
\centering
\includegraphics[width=0.6\linewidth]{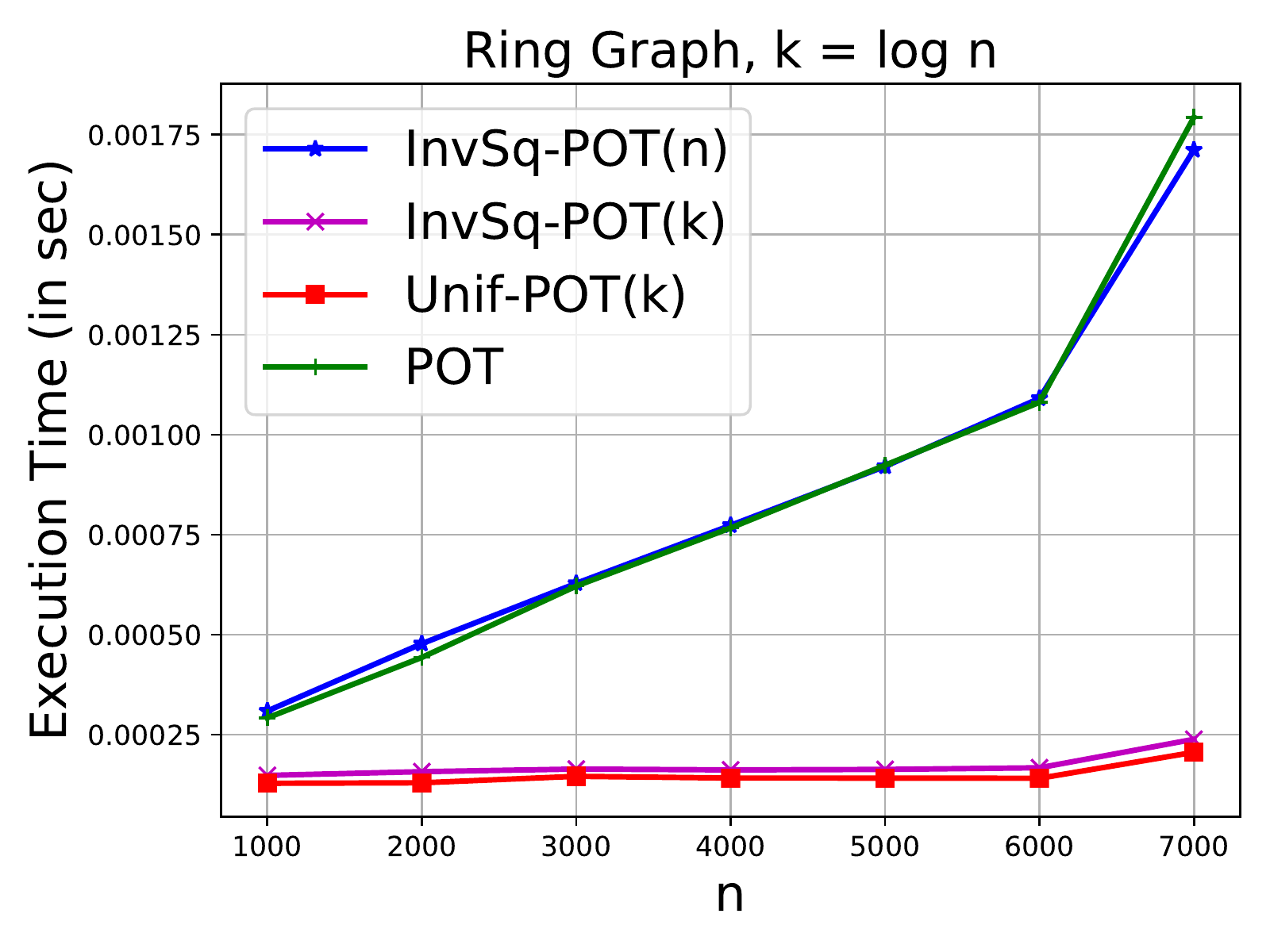}
\vspace{-0.1in}
\caption{Simulation Results for execution times of POT, Unif-POT($k$) and InvSq-POT($n$) with $k = \log n$ for Ring graph.}
\label{exec-times}
\vspace{-0.1in}
\end{figure} 

As mentioned in Section \ref{sec:model}, we evaluate the performance of proximity aware POT policies for three spatial graphs: Random Geometric (RG), Spatial Line (SL) and Spatial Ring (SR) graphs. We present the network parameters used for simulation in Table \ref{table:smspatial}. Note that the radius parameters for RG are chosen such that the graph remains almost surely connected asymptotically.

We first plot the total variation distance as a function of radial parameter $r$ of the RG topology as shown in Figure \ref{spatial} (a). Again to our surprise, for all values of $k =\log n, n$, both the proximity based policies produce a variation distance as low as $0.6\%.$ Note that for SL and SR topologies, we adopt a different job arrival model  to incorporate the spatial nature of job request pattern as discussed in Section \ref{sec:model}. To be precise we assume both jobs and servers are placed uniformly at random on a one dimensional line $[0, L_{max})$ and on a circle of radius $R$ for SL and SR topologies respectively. We plot variation distance as a function of network size for SL and SR as shown in Figures \ref{spatial} (b), (c). We observe a clear trend of Unif-POT($\log n$) and InvSq-POT($\log n$)  policy producing the smallest and largest variation distances for both SL and SR with InvSq-POT($\log n$) producing a variation distance of around $8\%.$ Also, note that these variation distances are insensitive to network size. Another thing to note that with introduction of spatial aspect, the variation distances increased by five folds as compared to their non-spatial counterparts (Figures \ref{fig:deter} (b) and (e)) for the same network size.

We next plot average request distance as a function of radial parameter of RG topology as shown in Figure \ref{spatial} (d). First note that the proximity aware policies are almost insensitive to $r.$ With increase in $r,$ the graph density for RG increases there by reducing average path length of the network. Thus we observe a decrease in average request distance for POT policy with increase in $r.$ As observed before, InvSq-POT($k$) produces lower request distances as compared to their Unif-POT($k$) counterpart. Lastly, InvSq-POT($k$) produces the least average request distance for SL which is almost insensitive to system size as shown in Figure \ref{spatial} (e). However, as expected, POT produces a very high request distance which scales linearly with system size.

\subsection{Evolution of Total Variation Distance}
Until now we have considered systems where there are an equal number of servers and jobs, i.e. when $m = n$. In this Section we analyze the system when $m < n$ or $m>n$ as we increase $m$ while keeping $n$ fixed. We design such an experiment to study the evolution of the total variation distance for a fixed $n$ and the underlying network topology. We evaluate the load distribution of POT policy and other proximity aware policies after each job arrival. We then calculate the total variation distance after each job arrival and plot its evolution as shown in Figures \ref{evolution} (a), (b) and (c). While we only focus on InvSq-POT($n$) policy for Figures \ref{evolution} (a)-(c), we get similar results for other proximity aware policies.

We consider a network of $n = 5000$ servers. We observe the system from arrival of first job till the $10000^{th}$ job. We plot the evolution of variation distance for various random network topologies as shown in Figure \ref{evolution} (a). We set the parameters for all three random graphs: ER, BA and RR such that their graph density remains almost equal. To be precise we set $\alpha = \log n, \beta = 2\log n$ and $\gamma = 2\log n/n.$ First we observe that with increase in number of jobs the total variation distance continuously  increases for RR and ER networks. However, for this particular choice of network parameters, variation distance for BA first increases and then decreases. Thus one can believe that proximity aware policies on scale free networks may provide good load balancing properties when there is imbalance between the number of servers and number of jobs. However, no such phenomena is observed for non-spatial and spatial Line and Ring topologies.

\subsection{Comparison of Execution Times}
POT policy samples a server uniformly at random from the remaining $n-1$ servers while a UnifPOT($k$) policy uniformly samples from a smaller number of candidate servers. Similarly, non-uniform sampling is involved when implementing InvSq-POT($k$) policy. Thus algorithmically, the execution time for each of the policy depends on the sample size ($n$ vs $k$) of the candidate set of servers and the nature of the sampling (uniform vs non-uniform). 

We plot the computation time for different proximity aware policies along with POT as a function of number of servers as shown in Figure \ref{exec-times}. We calculate the execution time of these policies for a random job allocation averaged over $50$ runs. We assume the distance information for InvSq-POT($k$) policy is known and precomputed beforehand. We assume the servers are connected through a Ring topology. It is clear from Figure \ref{exec-times} that the performance of POT and InvSq-POT($n$) are almost identical due to similar sampling set size. However, the execution times for InvSq-POT($k$) and  UnifPOT($k$) is far less due to a smaller sampling set size. The execution time for InvSq-POT($k$) is slightly larger than that of UnifPOT($k$)  due to non-uniform nature of sampling in the former.

\section{Simulation Results: Dynamic Load Balancing Systems}\label{sec:dyn}
Under the static load balancing model, the proposed proximity based policies have achieved a load distribution very similar to that of POT policy while reducing the average request distance to a great extent.  We now study these policies both in terms of performance and the communication cost for dynamic load balancing systems. We use the following parameters in our simulations: $n=1001$, $\lambda=0.95$, $\mu=1$, $k\in\{2,3,5,10,15,20,125,500\}$, and $d=2$. Each simulation consists of $n\times 10^5$ job arrivals.
\subsection{Comparison of PDFs of InvSq-POT(k) and Unif-POT(k) Policies}


\begin{figure*}[!htbp]
	\centering
	\begin{minipage}{0.5\textwidth}
		\includegraphics[width=1\textwidth]{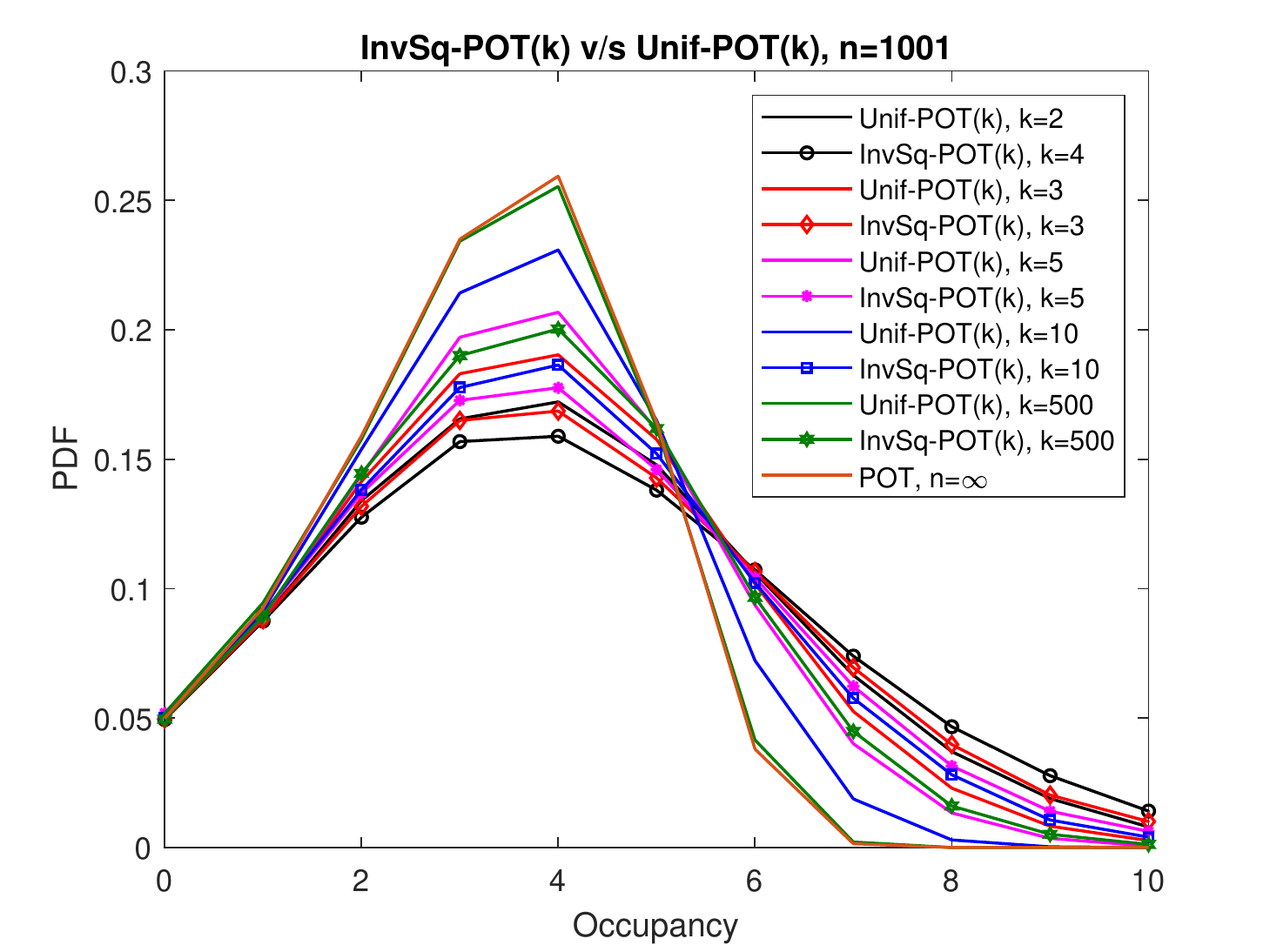}
	\end{minipage}
	\caption{The stationary distribution of a server for InvSq-POT(k) and Unif-POT(k)}
	\label{fig:inv_podk_v1_n=1001}
\end{figure*}


In Figure~\ref{fig:inv_podk_v1_n=1001}, we compare the resulting PDFs when we use InvSq-POT(k) and Unif-POT(k) policies for different values of $k$.


In Figure~\ref{fig:inv_podk_v1_n=1001}, we consider the case when $k\in\{2,3,5,10, 500\}$. We also plot the fixed-point of the mean-field limit (POT, $n=\infty$) for comparison purposes.
From Figure~\ref{fig:inv_podk_v1_n=1001}, we observe that the PDFs are very different for InvSq-POT(k) and Unif-POT(k) policies for different values of $k$. For $k=1$, both policies give the same performance. However, the gap between the PDFs for the same value of $k$ increases with $k$. 
This might be due to the fact that the distance between the sampling probability vectors under two policies increases with $k$.


\subsection{Comparison of PDFs of InvSq-POT(k) and POT Policies }

\begin{figure*}[!htbp]
	\centering
	\begin{minipage}{0.5\textwidth}
		\includegraphics[width=1\textwidth]{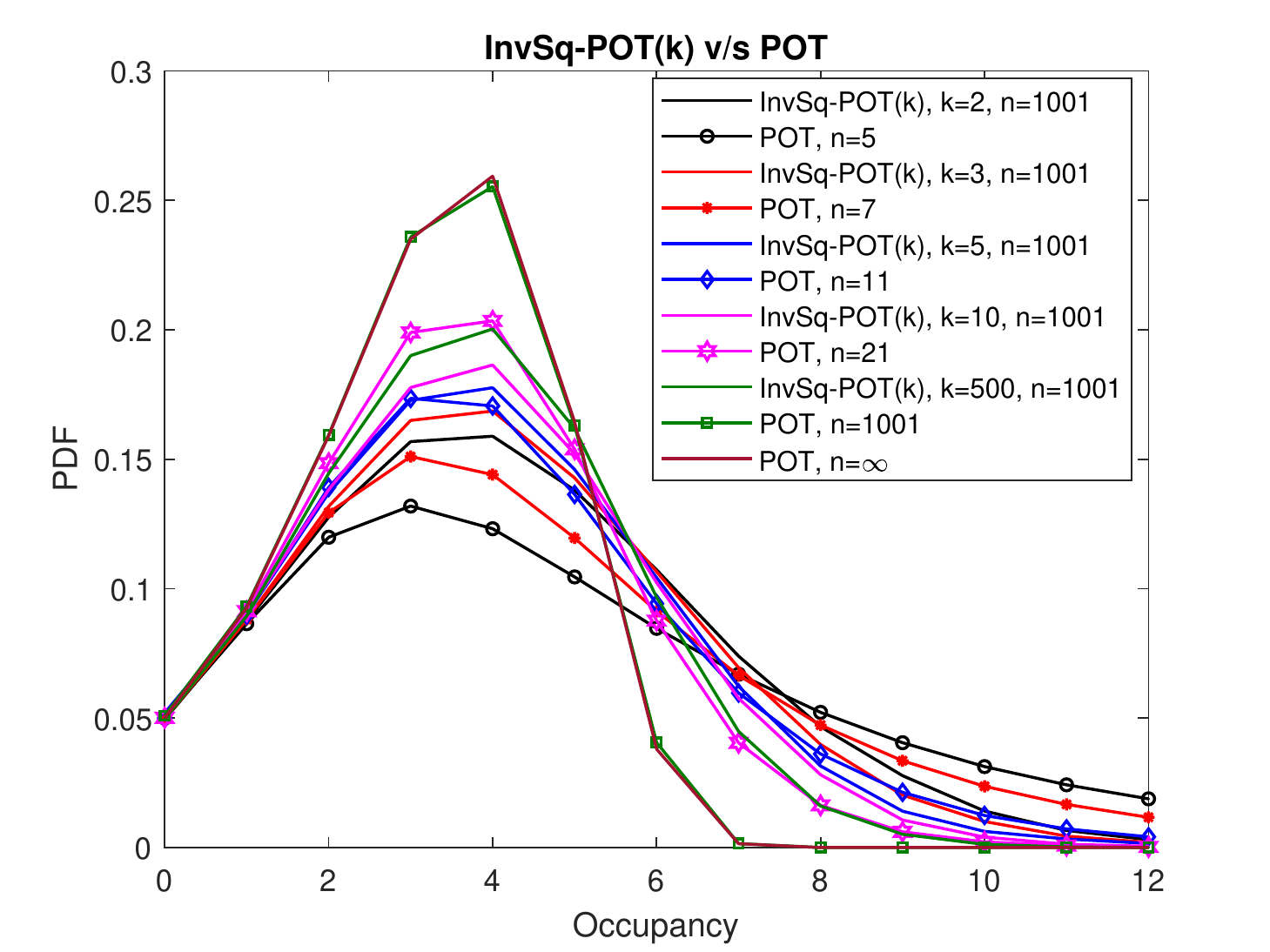}
		\subcaption{}
	\end{minipage}
	\begin{minipage}{0.5\textwidth}
		\includegraphics[width=1\textwidth]{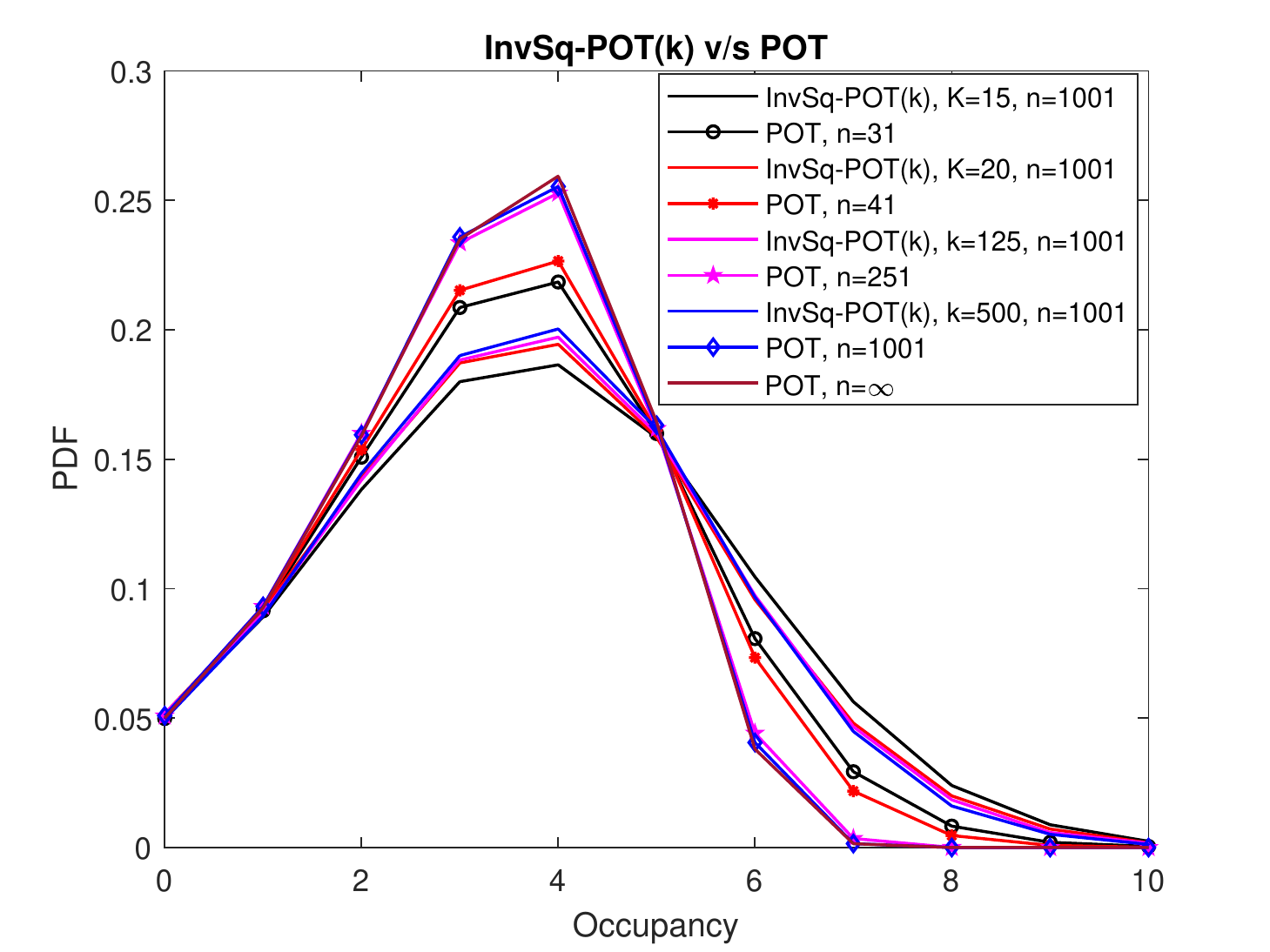}
		\subcaption{}
	\end{minipage}
	\caption{The stationary distribution of a server for InvSq-POT(k) and POT}
	\label{fig:inv_pod_v1_n1001}
\end{figure*}

In this section, we compare the performance of InvSq-POT(k) policy with that of POT policy for the system with $n=2k+1$. From Figures~\ref{fig:inv_pod_v1_n1001} (a) and \ref{fig:inv_pod_v1_n1001} (b), we observe that there is a significant mismatch between the resulting PDFs for InvSq-POT(k) and POT policies for fixed $k$. Also, from Figure~\ref{fig:inv_pod_v1_n1001} (a), we observe that the PDF for InvSq-POT(k) policy with $k=500$ is very close to the PDF for the POT policy for the system with $n=21$. This shows underperformance of InvSq-POT(k) policy.
%
\subsection{Comparison of PDFs for Different Values of $n$}
\begin{figure*}[!htbp]
	\centering
	\begin{minipage}{0.5\textwidth}
		\includegraphics[width=1\textwidth]{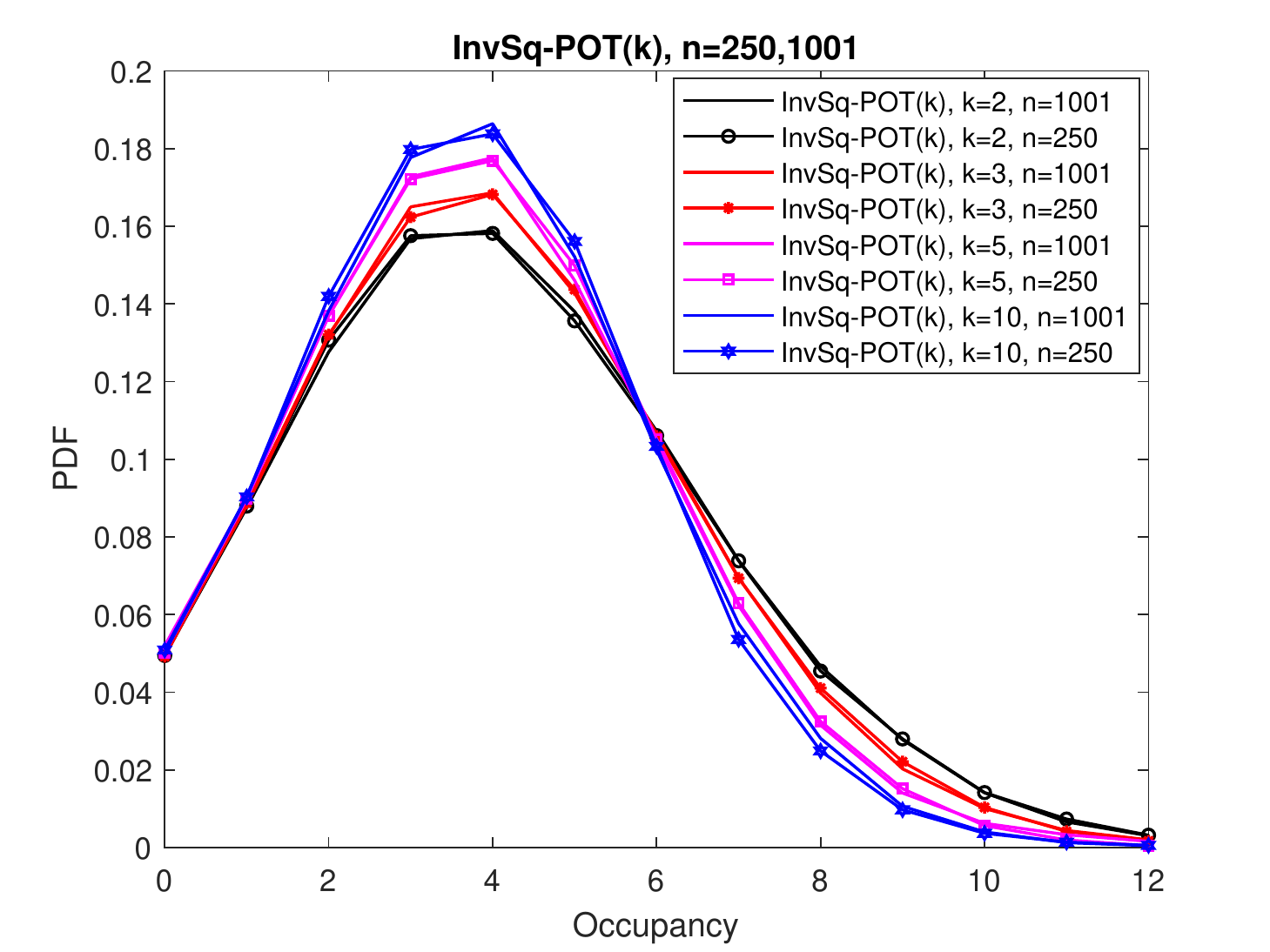}
		\subcaption{}
	\end{minipage}
	\begin{minipage}{0.5\textwidth}
		\includegraphics[width=1\textwidth]{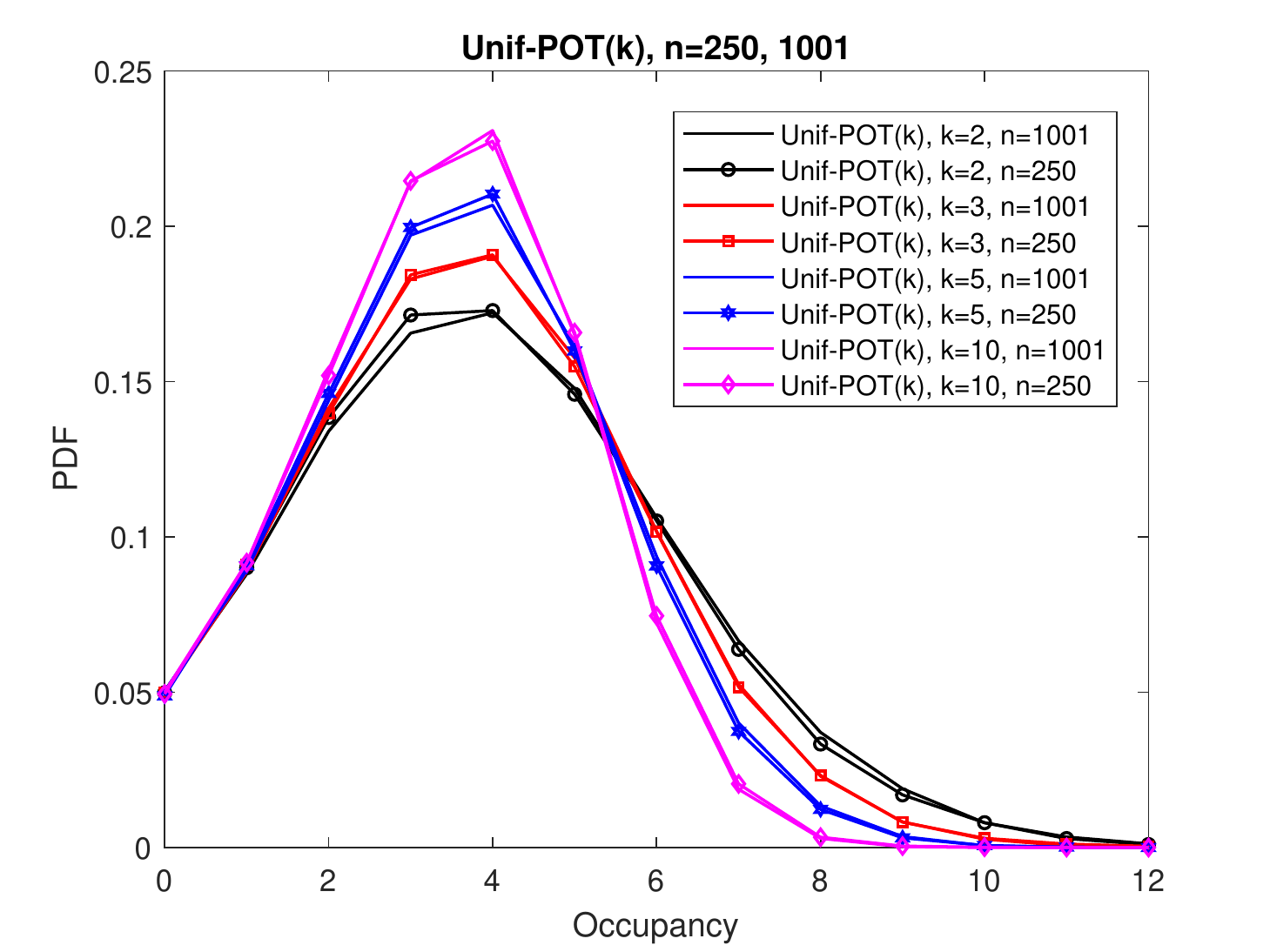}
		\subcaption{}
	\end{minipage}
	\caption{The stationary distribution of a server for (a) InvSq-POT(k) policy and (b) Unif-POT(k) policy for $n=250$ and $n=1001$}
	\label{fig:varying_N}
\end{figure*}
In this section, we compare the PDFs of InvSq-POT(k) and Unif-POT(k) policies for $n=250$ and $n=1001$ when $k\in\{2,3,5,10\}$. From Figures~\ref{fig:varying_N} (a) and \ref{fig:varying_N} (b), we observe that the PDFs almost coincide for $n=250$ and $n=1001$. This suggests that as $n\to\infty$, there exists a limiting PDF for a server's state.

\subsection{Asymptotic analysis when $n\to\infty$ for fixed $k$}

\begin{figure*}[!htbp]
	\centering
	\begin{minipage}{0.5\textwidth}
		\includegraphics[width=1\textwidth]{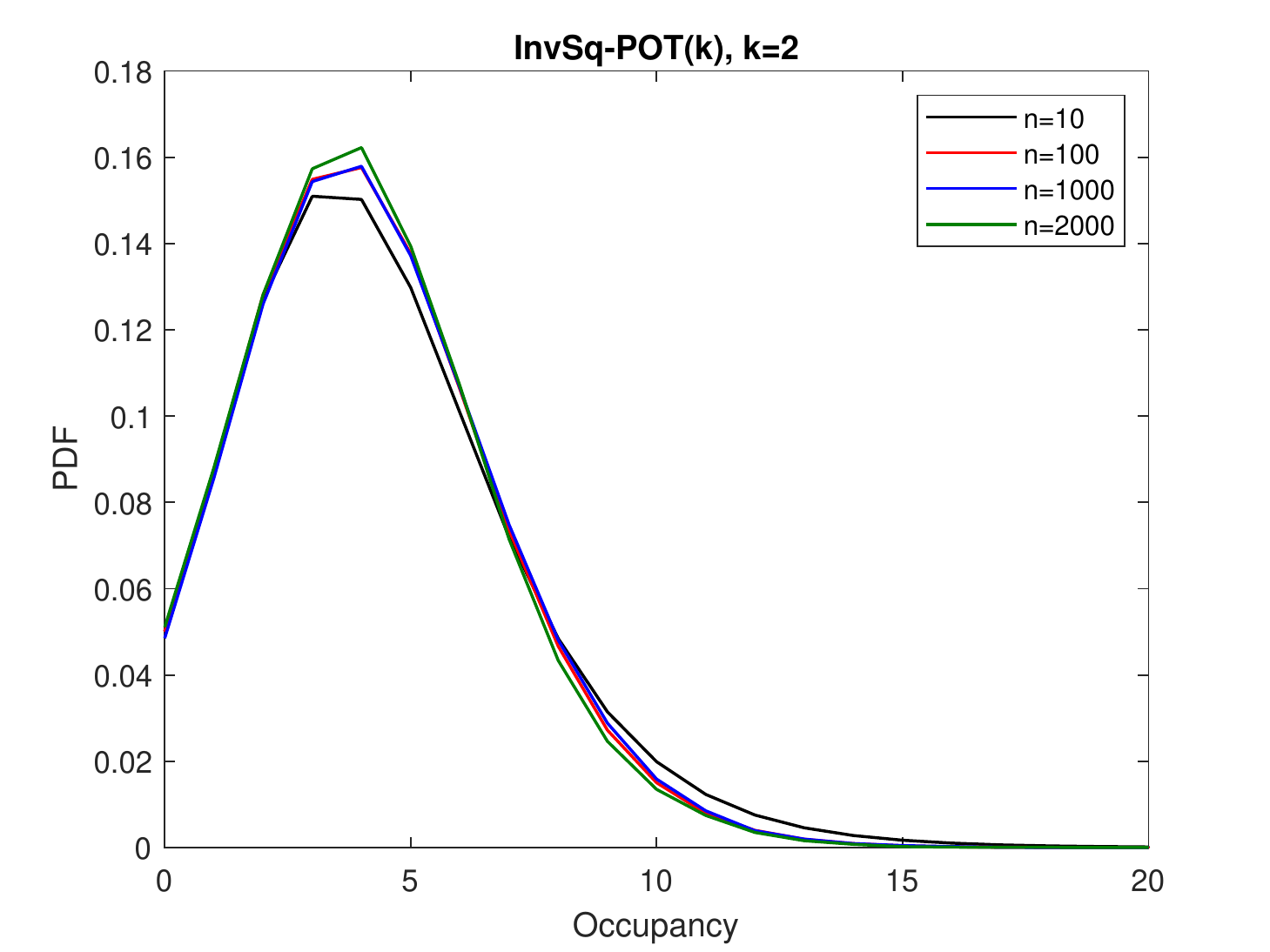}
		\subcaption{}
	\end{minipage}
	\begin{minipage}{0.5\textwidth}
		\includegraphics[width=1\textwidth]{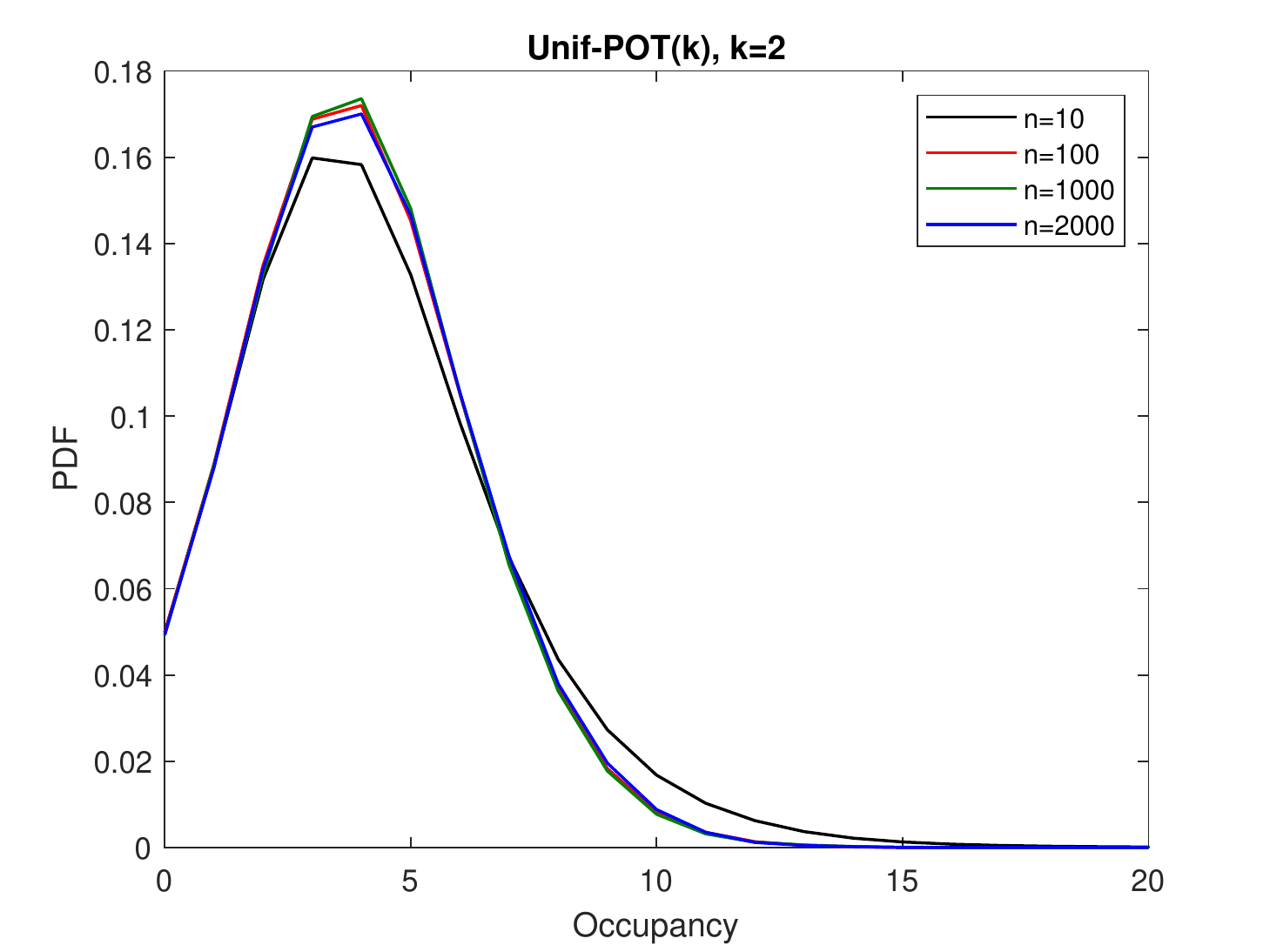}
		\subcaption{}
	\end{minipage}
	\caption{The stationary distribution of a server for (a) InvSq-POT(k) policy and (b) Unif-POT(k) policy for different values of $n$}
	\label{fig:k4_varying_N_v2}
\end{figure*}

Let $Q_i$ be the state of server $i$ and $P_t(Q_1=m_1,\cdots,Q_n=m_n)$ is the joint probability Server $i$ has $m_i$ jobs at time $t$, $1\leq i\leq n$. For $\ul{m}=(m_1,\cdots,m_n)$, let $\bm{\pi}^{(k,n)}=(\pi^{(k,n)}(\ul{m}),m_i\in\mb{Z}_+,1\leq i\leq n)$, where $\pi^{(k,n)}(\ul{m})$ is the joint stationary probability that Server $i$ has $m_i$ jobs for all $1\leq i\leq n$. Let $\bm{\alpha}^{(k,n)}=(\alpha^{(k,n)}(c),c\in\mb{Z}_+)$ be the stationary distribution of a server defined as
\beq
\alpha^{(k,n)}(m_1)=\sum_{m_2,\cdots,m_n}\pi^{(k,n)}(\ul{m}).
\eeq

In Figure~\ref{fig:k4_varying_N_v2}, we plot PDFs for $k=2$, $\lambda=0.95$, and $n\in\{10,100,1000,2000\}$. From Figures~\ref{fig:k4_varying_N_v2} (a) and \ref{fig:k4_varying_N_v2} (b), we observe that there exists a distribution $\bm{\alpha}^{(k)}=(\alpha^{(k)}(c),c\in\mb{Z}_+)$ satisfying
\beq
\lim_{n\to\infty}\bm{\alpha}^{(k,n)}=\bm{\alpha}^{(k)},
\eeq
where the limit is in distribution sense.

Next, we observe that the joint PDF of two servers depends on the distance between these two servers. Let $\alpha^{(k,n)}(Q_i=m_i,Q_j=m_j)$ is the joint probability that server $i$ has $m_i$ jobs and server $j$ has $m_j$ jobs. Then we define
\beq
\norm{\alpha^{(k,n)}(Q_i,Q_j)-\alpha^{(k,n)}(Q_l,Q_m)}=\frac{1}{2}\sum_{u,v}\abs{\alpha^{(k,n)}(Q_i=u,Q_j=v)-\alpha^{(k,n)}(Q_l=u,Q_m=v)}.
\eeq
We compute $\alpha^{(k,n)}(Q_1,Q_2)$, $\alpha^{(k,n)}(Q_1,Q_3)$, and $\alpha^{(k,n)}(Q_1,Q_9)$. From Tables~\ref{table:table4} and \ref{table:table5}, we observe that the joint PDF of two servers depends both on the distance between the two servers and how we choose the distance based sampling probabilities. For example, for Unif-POT(k) and InvSq-POT(k) policies, the distance between joint PDFs do not coincide indicating that the sampling probabilities that we choose affect the joint PDFs. For $n=10$, servers $3$ and $9$ are at equal distance from server $1$, as a result, the joint PDFs 
$\alpha^{(k,n)}(Q_1,Q_3)$ and $\alpha^{(k,n)}(Q_1,Q_9)$ must coincide due to symmetry of the system. This is also observed in Tables~\ref{table:table4} and \ref{table:table5}. Since both load balancing policies take location of servers into account while assigning a job, there is no mean-field effect. Otherwise the distances between different joint PDFs could have been very close to  zero for $n=1000$  and $n=2000$ due to propagation of chaos in the stationary regime.

\begin{table}
	\begin{center}
		\scalebox{0.9}{
			\begin{tabular}{ |c|c|c|c| } 
				\hline
				n & $\norm{\alpha^{(k,n)}(Q_1,Q_2)-\alpha^{(k,n)}(Q_1,Q_3)}$& $\norm{\alpha^{(k,n)}(Q_1,Q_2)-\alpha^{(k,n)}(Q_1,Q_9)}$& $\norm{\alpha^{(k,n)}(Q_1,Q_3)-\alpha^{(k,n)}(Q_1,Q_9)}$\\ 
				\hline
				10 & 0.0222& 0.0220&0.0013 \\ 
				1000 & 0.0303& 0.2150&0.1867 \\
				2000 & 0.0287&0.2105&0.1853\\
				\hline
			\end{tabular}
		}
		\caption{Properties of joint PDF for Unif-POT(k) Policy}
		\label{table:table4}
	\end{center}
\end{table}

\begin{table}
	\begin{center}
		\scalebox{0.9}{
			\begin{tabular}{ |c|c|c|c| } 
				\hline
				n & $\norm{\alpha^{(k,n)}(Q_1,Q_2)-\alpha^{(k,n)}(Q_1,Q_3)}$& $\norm{\alpha^{(k,n)}(Q_1,Q_2)-\alpha^{(k,n)}(Q_1,Q_9)}$& $\norm{\alpha^{(k,n)}(Q_1,Q_3)-\alpha^{(k,n)}(Q_1,Q_9)}$\\ 
				\hline
				10 & 0.1057& 0.1060&0.0013 \\ 
				1000 & 0.1127& 0.2889&0.1859 \\
				2000 & 0.1065&0.2803&0.1812\\
				\hline
			\end{tabular}
		}
		\caption{Properties of joint PDF for InvSq-POT(k) Policy}
		\label{table:table5}
	\end{center}
\end{table}

\subsection{The Average Weighting Time}
In Table~\ref{table:table4}, we compare the average weighting time of a job between InvSq-POT($k$) policy and Unif-POT($k$) policy for $n=1001$, $\lambda=0.95$ and $\mu=1.$ For $k=500$, Invsq-POT($k$) policy results in $13.44\%$ increment in the average weighting time over the POT policy.

\subsection{The Average Request Distance}
In Table~\ref{table:table3}, we compare the average request distance between the location where a job has arrived and the destination server. Clearly, InvSq-POT(k) policy outperforms Unif-POT(k) policy.

\begin{table}
	\begin{center}
		\begin{tabular}{ |c|c|c| } 
			\hline
			k & Unif-POT($k$)& InvSq-POT($k$)\\ 
			\hline
			2 & 4.1995& 4.4258 \\ 
			3 & 3.9417& 4.2671 \\
			5 & 3.7788&4.1038\\
			10& 3.5594&4.0091\\ 
			15& 3.5249&3.9698\\
			20& 3.4865 &3.8645\\
			125&3.3823&3.8535\\
			500&3.3801&3.8346\\
			\hline
		\end{tabular}
		\caption{The average weighting time of a job}
		\label{table:table4}
		\vspace{-6ex}
	\end{center}
\end{table}
\begin{table}
	\begin{center}
		\begin{tabular}{ |c|c|c| } 
			\hline
			k & Unif-POT(k)& InvSq-POT(k)\\ 
			\hline
			2 & 0.7499& 0.6 \\ 
			3 & 0.9999& 0.6736 \\
			5 & 1.4999&0.7801\\
			10& 2.7501&0.9452\\ 
			15& 4.0001&1.0499\\
			20& 5.2498 &1.1273\\
			125&31.4993&1.6531\\
			500&125.2450&2.0689\\
			\hline
		\end{tabular}
		\caption{The average request distance of a job}
		\label{table:table3}
		\vspace{-4ex}
	\end{center}
\end{table}

\subsection{Open Problems and Conjectures}
In this section, we discuss some problems that are of interest to study.
\subsection*{Problem~1:}
For given $k$, based on Figures~\ref{fig:varying_N} (a) and \ref{fig:varying_N} (b), it is of interest to obtain approximations for PDFs when $n$ is large. Mean-field techniques are not applicable in this case.

\subsection*{Problem~2}
We need to study the impact of topology on the performance. For highly connected graphs, the drop in performance for distance based sampling policies over the POT policy may not be significant, but we might also see a drop in the improvement in the average cost.


\section{Conclusion}\label{sec:con}
In this work we considered a class of proximity aware Power of Two choice policies for the case when servers are interconnected as an arbitrary graph. We evaluated the performance of these policies in both static and dynamic load balancing systems. We performed extensive simulations over a wide range of network topologies for the static systems. To our surprise, with few simple modifications in the server sampling process, we observed a drastic reduction in the overall system wide communication cost while a similar load distribution profile as that of POT policy. More precisely, we observe that a communication cost aware non-uniform sampling based POT policy: InvSq-POT($k$) achieves the best of both the worlds in terms of communication cost and load balancing. For dynamic load balancing systems over a ring topology, we observed a drop in performance while getting a significant drop in communication cost associated with both proximity based policies. Finally, going further, we aim at extending our simulation results to consider different graph topologies for the dynamic system.

\section{Acknowledgment}\label{ack}
This research was sponsored by the U.S. Army Research Laboratory and the U.K. Defence Science and Technology Laboratory under Agreement Number W911NF-16-3-0001 and by the NSF under grant NSF CNS-1617437. The views and conclusions contained in this document are those of the authors and should not be interpreted as representing the official policies, either expressed or implied, of the U.S. Army Research Laboratory, the U.S. Government, the U.K. Defence Science and Technology Laboratory. This document does not contain technology or technical data controlled under either the U.S. International Traffic in Arms Regulations or the U.S. Export Administration Regulations.


\bibliographystyle{ACM-Reference-Format}
\bibliography{refs} 

\end{document}